\documentclass[sort&compress
    ,final            
  ]
  {aipproc}

\usepackage{sidecap,graphicx,boxedminipage,epsfig,wrapfig,floatflt,shadow}
\layoutstyle{8x11double}
\pdfoutput=1

\begin{document}

\title{Surveying Graduate Students' Attitudes and Approaches to Problem Solving}

\author{Andrew Mason and Chandralekha Singh}{
  address={Department of Physics, University of Pittsburgh, Pittsburgh, PA, 15213}
}



\begin{abstract}
Students' attitudes and approaches to problem solving in physics can profoundly influence their motivation to learn and development of 
expertise. We developed and validated an Attitudes and Approaches to Problem Solving survey by expanding the Attitudes towards Problem Solving
survey of Marx and Cummings~\cite{cummings2} and administered it to physics graduate students.
Comparison of their responses to the survey questions about problem solving in their own graduate level courses vs. problem solving in the 
introductory physics courses provides insight into their expertise in introductory and graduate level physics. 
The physics graduate students' responses to the survey questions were also compared with those of introductory 
physics and astronomy students and physics faculty. We find that, even for problem solving in introductory physics,
graduate students' responses to some survey questions are 
less expert-like than those of the physics faculty. Comparison of survey responses of graduate students and introductory students 
for problem solving in introductory physics suggests that graduate students' responses are in general more expert-like than
those of introductory students. However, survey responses suggest that graduate-level problem solving by graduate students 
on several measures has remarkably similar trends to introductory-level problem solving by introductory students. 

\end{abstract}

\maketitle
\section{Introduction}

Students' attitudes and approaches towards learning can have a significant impact on what students actually 
learn.~\cite{hammer3,schommer,vass,mpex,class1,class2} Mastering physics amounts to not only developing a robust knowledge structure
of physics concepts but also developing productive attitudes about the knowledge and learning in physics. In essence, it is impossible to
become a true physics expert without a simultaneous evolution of expert-like attitudes about the knowledge and learning in physics.
If students think that physics is a collection of disconnected facts and formulas rather than
seeing the coherent structure of the knowledge in physics, they are unlikely to see the need for organizing their
knowledge hierarchically. Similarly, if students believe that only a few smart people can do physics, the teacher
is the authority and the students' task in a physics course is to take notes, memorize the content and reproduce it on the exam and then forget it,
they are unlikely to make an effort to synthesize and analyze what is taught, ask questions about how concepts fit together or
how they can extend their knowledge beyond what is taught. Similarly, if students believe that if they cannot solve a problem within 10 
minutes, they should give up, they are unlikely to persevere and make an effort to explore strategies for solving challenging problems. 

The Maryland Physics Expectation Survey (MPEX) was developed to explore students' attitudes and expectations related to physics.~\cite{mpex} 
When the survey was administered before and after instruction in various introductory physics courses, it was found that students'
attitudes about physics after instruction deteriorated compared to their expectations before taking introductory physics. Very few carefully
designed courses and curricula have shown major improvements in students' expectations after an introductory physics 
course.~\cite{cummings2,elby,laws}

Colorado Attitudes about Science Survey (CLASS) is another survey which is similar to the MPEX survey and explores students' attitudes about 
physics.~\cite{class1,class2}
The analysis of CLASS data yields qualitatively similar results to those obtained using the MPEX survey. 
Moreover, when introductory physics students were asked to answer the survey questions twice, once providing
the answers from their perspective and then from the perspective of their professors,
introductory students' responses to many questions were very different from their perspective compared to what they claimed would
be their professors' perspective.~\cite{class2} 
Thus, introductory students maintained their views although they knew that the physics professors would have different views about some of the survey questions.

Cummings et al.~\cite{cummings2,cummings1} developed an Attitudes towards Problem Solving Survey (APSS) which is
partially based upon the Maryland Physics Expectations Survey (MPEX). The original APSS survey
has 20 questions and examines students' attitudes towards physics problem solving.~\cite{cummings1}  
The survey was given to students before and after instruction at three types of institutions: a large university, a smaller university 
and a college. It was found that students' attitudes about problem solving did not improve after instruction (deteriorated slightly) at the
large University and the attitudes were least expert-like (least favorable) at the large University with a large class size.

Students' attitudes and approaches to learning and problem solving can affect how they learn and how much time they spend repairing, 
extending and organizing 
their knowledge structure. If instructors are aware of students attitudes and approaches to problem solving, they can explicitly exploit strategies to
improve them. For example, knowing students' beliefs about mathematics learning (which is similar to students' beliefs about physics learning
in many aspects) motivated Schoenfeld to develop a curriculum to improve students' 
attitude.~\cite{cog1,schoenfeld1,schoenfeld2,schoenfeld3}
In particular, based on the knowledge that
students in the introductory mathematics courses often start looking for formulas right away while solving a mathematics problem instead
of performing a careful conceptual analysis and planning, Schoenfeld used an explicit strategy to change students' approach. He routinely placed
students in small groups and asked them to solve problems. He would move around
and ask them questions such as ``What are you doing? Why are you doing it? How does it take you closer to your goals?"
Very soon, students who were used to immediately looking for formulas were embarrassed and realized that they should
first perform conceptual analysis and planning before jumping into the implementation of the problem solution. Schoenfeld's strategy helped
most students adopt an effective problem solving approach within a few weeks and they started to devote time to qualitative analysis 
and decision making before looking for equations.~\cite{cog1}

Another unfavorable attitude about mathematical problem solving that Schoenfeld wanted students to give up was
that students often felt that if they could not solve a problem within 5-10 minutes, they should give 
up.~\cite{cog1,schoenfeld1,schoenfeld2,schoenfeld3} Schoenfeld realized
that one reason students had such an attitude was because they saw their instructor solving problems during the lectures without
faltering or spending too much time thinking. To bust this myth about problem solving, Schoenfeld began each of his geometry classes with
the first 10 minutes devoted to taking students' questions about challenging geometry problems (often from the end of the chapter excercises)
and thus attempting to solve them without prior preparation. Students discovered that Schoenfeld often struggled with the problems and was 
unable to solve them in the first 10 minutes and asked students to continue to think about the problems until one of them had solved it
and shared it with others. This approach improved students' attitude and their self-confidence in solving mathematics problems.

Here, we briefly discuss the development, validation and administration of the Attitudes and Approaches to Problem Solving (AAPS) survey, a modified version of 
APSS survey~\cite{cummings1}, that includes additional questions related to approaches to problem solving and we focus on the responses of physics graduate students.~\cite{perc}
We explore how graduate students differ in their attitudes and approaches while they solve graduate level problems versus introductory level problems. 
The survey questions were administered in the form of statements that one could agree 
or disagree with on a scale of 1 (strongly agree) to 5 (strongly disagree) with 3 signifying a neutral response.  
We do not differentiate between ``agree" and ``strongly agree" in interpreting the data (both are coded as +1). Similarly, ``disagree" and ``strongly
disagree" were combined (both are coded as -1) for streamlining the data and their interpretation. 
A favorable response refers to either ``agree" or ``disagree" based upon which one was favored by a majority of physics faculty. 
We find that, on some measures, graduate students have very different attitudes and approaches about solving 
introductory physics problems compared to their own graduate level problems.
The attitudes and approaches of graduate students on the AAPS survey was also compared to those of introductory physics and 
astronomy students and to physics faculty. We find that the attitudes and approaches of graduate students differ significantly from introductory 
students and physics faculty on several measures.  

\section{The Groups that were Administered the AAPS Survey}

The final version of the AAPS survey was first administered 
anonymously to 16 physics graduate students enrolled in a graduate level TA training course at the end of the one-semester course. 
The expert (favorable) responses are given in the Appendix along with the survey.
Discussions with the graduate students after they took the survey showed that all of them interpreted that the survey
was asking about problem solving in their own graduate courses and that they would have answered the questions
differently if they were asked about their attitudes and approaches to solving introductory physics problems.
Then, we administered the survey to 24 graduate students (there was overlap between the first cohort of 16 graduate students
and this cohort) with the questions explicitly asking
them to answer each question about their attitudes and approaches to introductory physics problem solving.
Due to lack of class time, this second round of survey was administered online. 
We had individual discussions with 4 graduate students about the reasoning for their AAPS survey responses and invited all 24 graduate students 
who had answered the questions online to write a few sentences explaining their reasoning for selected survey questions online. We explicitly
asked them to explain their reasoning when they answered the survey questions about problem solving in the graduate level courses
and separately for introductory physics. Ten graduate students (out of 24 who took the survey online) provided written reasonings for their responses.
The following year, the survey was administered to 18 graduate students at the end of a graduate level TA training course in which the graduate students
were first asked to respond to the survey questions for problem solving in their own graduate courses and then for introductory physics problem solving (after they
had submitted their survey responses to the graduate-level problem solving).

We also administered the AAPS survey to several hundred introductory students in two different first-semester and second-semester algebra-based physics 
courses and to students in the first and second-semester calculus-based courses. In particular, there were two sections of the first-semester algebra-based physics
course with 209 students, two sections of the second-semester algebra-based physics course with 188 students,  one first-semester calculus-based
course section with 100 students and a second-semester calculus-based course section with 44 students. In all of these courses, students were given small
amounts of bonus points for taking the survey.  In addition, the survey was given to 31 students in an astronomy course which 
is the first astronomy course taken by students who plan to major in Physics and Astronomy (but less than $20\%$ of the students in this
course actually end up majoring in Physics and Astronomy). None of the students in the 
astronomy course, who did not want to major in physics and astronomy, were required to take that course unlike the students in
the introductory physics courses (the calculus-based introductory courses are dominated by engineering majors and algebra-based courses by
those interested in health related professions who must take two physics courses to fulfill their requirements). For the astronomy course, 
the word ``physics" in the survey everywhere was replaced by 
``astronomy", e.g., ``in solving astronomy problems...". Also, the contexts (which were not related to astronomy) were removed from the 
last two questions (32) and (33) of the survey for astronomy students.  Finally, the survey was given to 12 physics faculty who had taught introductory
physics recently. Half of the faculty members were those who also gave the survey to their introductory students. The faculty member answered the survey questions
for both introductory-level and graduate-level problem solving. 
We also discussed faculty responses to selected questions individually with some of them.

\section{Validity and Reliability of the Survey}

We now address issues related to validity and reliability of the survey. 
Reliability refers to the relative degree of consistency between the survey scores, e.g., if an individual repeats the procedures.~\cite{nitko} 
One measure of reliability of a survey is the Cronbach's alpha ($\alpha_c$) which establishes the survey's reliability via internal consistency check.  
The Cronbach's alpha ($\alpha_c$) test was applied over all 33 questions for all groups (N=672)
and the $\alpha_c=0.82$ which is reasonable from the standards of test-design.~\cite{nitko} As noted later, there is very little variability in the 
responses for some of the groups (e.g., faculty) so it does not make sense to calculate $\alpha_c$ separately for the various groups.

Validity refers to the appropriateness of interpreting the survey scores.~\cite{nitko} 
In order to develop the AAPS survey, we selected 16 questions from the APSS survey~\cite{cummings1}
and tweaked some of the questions for clarity based upon in-depth interviews with five introductory physics students 
and three physics faculty members.
These 16 questions constitute the first 14 questions and the last two questions of the AAPS survey, provided in the Appendix. 
We also developed 17 additional questions which were based upon discussions with the faculty members about productive approaches to problem solving, and 
modified them based upon the feedback from introductory students during interviews and discussions with some graduate students and three physics faculty members. 

Content validity refers to the degree to which the survey items reflect the domain of interest (in our case, attitudes and approaches to problem solving).~\cite{nitko}
As noted earlier, we discussed with some faculty members their opinions about productive approaches to problem solving and took their opinions into account
while developing the additional survey questions.
We further addressed the issue of content validity by taking measures to ensure that the respondents interpret the survey questions as was intended.  
To this end we interviewed sample respondents from the introductory course, physics graduate students (mostly those enrolled in a TA training course) and faculty members.  
During the interviews and discussions, we paid attention to respondents€™ interpretations of questions and modified
the questions accordingly in order to make clear the actual intent of the questions.  
While the interviews with the introductory students were formal and tape-recorded, the discussions with the faculty members and graduate students were informal and
were not tape-recorded.
The reason introductory physics students and faculty were sought for this purpose (in addition to the graduate students) is that we hypothesized that the responses of these
two groups would be the most disparate and would provide the most diverse set of feedback for improving the preliminary survey.
Some of the themes in the additional questions are related to the use of diagrams and scratch work in problem solving, 
use of ``gut" feeling vs. using physics principles to answer conceptual questions, reflection on one's solution after solving a problem to
learn from it, giving up on a problem after 10 minutes, preference for numerical vs. symbolic problems and enjoying solving challenging physics problems.
The in-depth interviews with five students from a first-semester algebra-based class, and discussions with the graduate students and three physics faculty members helped
modify the survey and were helpful in ensuring that the questions were interpreted clearly by both the experts and students at various anticipated levels of expertise.  

Of approximately 40 introductory students responding to the invitation for paid interviews from an introductory physics course, 
five were selected. Since we wanted all students to be able to interpret the problems, two students were randomly chosen for interview from those 
who scored above $70\%$ and three students were chosen who obtained below $70\%$ on their first midterm exam.
The survey questions were administered to all interviewed students in the form of statements that they could agree 
or disagree with on a scale of 1 (strongly agree) to 5 (strongly disagree) with 3 signifying a neutral response.  
During the individual interviews, students were also asked to solve some physics problems using a think-aloud protocol to gauge whether
what they answered in the survey questions about their attitudes and approaches to problem solving were consistent with
the attitudes and approaches actually displayed during problem solving. Within this protocol, we asked individuals to talk aloud while 
answering the questions. We did not disturb
them while they were talking and only asked for clarifications of the points they did not make clear on their own later. 
While it is impossible to grasp all facets of problem solving fully by having students
solve a few problems, a qualitative comparison of their answers to the survey questions and their actual approaches to solving problems
was done after the interviews using the think aloud protocol. This comparison suggests that students were consistent in their survey responses 
in many cases but in some instances they selected more favorable (expert-like) responses to the survey questions than the expertise that was 
explicitly visible from their actual problem solving.
In this sense, the favorable responses (at least for the introductory students) should be taken as the upper limit of the actual
favorable attitudes and approaches to problem solving.
      
We also tested validity of the survey by comparing actual survey data with those predicted according to the assumption of expert-novice behaviors, 
pre-defining the majority faculty response for each question as the 
``expert" response.~\cite{nitko}  In Table 1, we display data for individual questions for each statistical group as well as the 
average response for all 33 questions for each of the groups in the ``Avg." column.  
In the data reported in Table 1, the average score is as defined by Cummings et al.~\cite{cummings1} To calculate the 
average score for a question, a +1 is assigned to each favorable response, a -1 is assigned to each unfavorable response, and a 0 is assigned to neutral responses. 
We then average these values for everybody in a particular group (e.g., faculty) to obtain an average score for that group.
Thus, the average score on each question for a group indicates how expert-like the survey response of the group is on each survey question.
Table 1 shows that the faculty had unanimous or close to unanimous agreement on most of the survey questions. 
These results support content validity of the survey.
Table 1 also shows that faculty members answered the questions in a more expert-like fashion than graduate students, who in turn were more expert-like than introductory-level students.  
The difference between faculty and graduate students holds both for problem solving in graduate-level problems as well as problem solving
in introductory-level problems. Similarly, the graduate students' responses are more expert-like for graduate-level problems than introductory students' responses
are about problem solving at the introductory-level.  While these differences cannot be quantified apriori, such differences can be expected based upon the known expertise of each
of these groups in physics. 
All these differences are statistically significant ($p<0.05$).   
These consistencies further provide validity to the survey.

To determine whether the differences between the groups are statistically significant and there is an appreciable effect size, we examined the groups as follows: all introductory physics students 
(we combined these classes since we did not find statistical differences between different introductory physics classes); 
the astronomy group; all graduate students for introductory problems; all graduate students for graduate-level problems; all faculty for introductory problems; 
and all faculty for graduate-level problems.  
The effect sizes between groups over all 33 questions were calculated in the form of Cohen's means difference (= $(\mu_1-\mu_2)/\sigma_{pooled}$), 
calculating individual group means (on a scale of -1 to +1) and standard deviations.  Table 2 shows that the effect sizes between groups of different levels of 
expertise have a large to very large effect size ($1 < d < 2.5$), in favor of the more assumed expert-like group.~\cite{nitko}  
Individual p-values for pairwise t-tests~\cite{nitko} between each group shows that all differences are statistically significant except for the difference between the two faculty groups. 
Again these effect sizes are qualitatively consistent with the expected trends based upon the expertise of each group and provides validity to the survey.

\section{Results}

As noted earlier, the survey questions are designed 
in a ``agree" or ``disagree" format (see the Appendix). There is leeway to agree or disagree ``somewhat" or ``strongly" (although we did not distinguish between these in our
analysis presented here), and one may also select a neutral response. 
The favorable (expert) responses for each question based upon the responses chosen by most physics faculty are also given in the Appendix.
Table 1 shows the net average responses for all groups for each individual question and averaged over all questions (see 
the last entry in Table 1). 
In Table 1, the introductory physics students from both semesters of algebra-based and calculus-based courses 
were lumped into one group because there was no significant difference between the ``net" average survey responses of these groups.

In Table 1, we followed the method for reporting data used by Cummings et. al~\cite{cummings1}.
A second method for representing data separately shows the average percentage of favorable and unfavorable responses for each question for
each group (the neutral responses are $100\%$ minus the percentage of favorable and unfavorable responses). We will use this second method of data 
representation for all of our graphical representations of data below using histograms.
Moreover, since there are no significant differences between faculty responses for problem solving at the introductory and graduate levels except for a couple of questions (see Table 1), 
the histograms will only display one of the two sets (faculty responses for introductory level problem solving). 

We now examine graduate students' survey responses to individual questions (recall that the effect size between them and faculty in Table 2 appears to be similar to 
the effect size between them and introductory students). 
Comparison of the graduate students' problem solving attitudes and approaches for introductory and graduate level problems in Table 2 shows that there is small to moderate effect size ($0.4 < d < 0.6$).  
However, if problem solving attitudes and approaches of a given group (e.g., graduate students) did not depend at all on how difficult the problem was (e.g., whether it was introductory or
graduate level problem), there would be a very small effect size ($d < 0.2$).  
This moderate effect size suggests a difference in attitudes and approaches for a group with different levels of problems (introductory vs. graduate level), 
and is discussed in the following sections.  

\subsection{Comparison of graduate students' survey responses for graduate level vs. introductory level problem solving}

Figure 1 compares the AAPS survey responses of graduate students to selected questions for which differences were observed
when they answered the questions about problem solving in their graduate courses and problem solving in introductory physics.
The error bars on the histograms (and in all the other figures in this paper) show the standard error.
One typical difference between introductory and graduate level problem solving is that the graduate students 
display more expert-like (favorable) attitudes and approaches 
while solving introductory level problems than while solving graduate level problems.
For example, in response to question (1) for problem solving in their graduate level courses, approximately $40\%$ of the graduate students 
felt that if they were not sure about the right way to start a problem, they would be stuck unless they got help but only $20\%$ felt this 
way when solving introductory physics problems. 
Also, they were more likely to reflect upon physics principles that may
apply and see if they yield a reasonable solution when not sure about the approach while solving introductory problems than while
solving graduate level problems (see response to question (10) in Figure 1). 
They were also more likely to be able to tell that their answer was wrong without external
input while solving introductory problems than graduate level problems (see response to question (6) in Figure 1). 
Graduate students were approximately $20\%$ more likely to claim that they routinely use equations to calculate answers even if they
are non-intuitive while solving graduate level problems than while solving introductory level problems (see response to question (11) in Figure 1).

While none of the graduate students claimed they would give up solving an introductory physics problem if they could not solve it
within 10 minutes, approximately $15\%$ claimed they would give up after 10 minutes while solving a graduate level 
problem (see response to question (23) in Figure 1).
Moreover, while approximately $80\%$ of the graduate students claimed they enjoy solving introductory physics problems even though it can be challenging
at times, less than $70\%$ of them said the same about the graduate level problems (see response to question (27) in Figure 1).
Also, more graduate students claimed that 
it is useful for them to solve a few difficult problems using a systematic approach and learn from 
them rather than solving many similar easy problems one after another when solving introductory level problems than for 
graduate level problems (see response to question (26) in Figure 1).

As shown in Table 1, the introductory physics students noted that they enjoyed solving challenging problems even less than the graduate students and were also
less likely to find solving a few difficult problems more useful than solving many easy problems based upon the same principle 
(see introductory students' responses to question (26) and (27) in Table 1).
One introductory student stated in an interview that he feels frustrated with his incorrect problem solution
and feels satisfied when he gets a problem right, which motivates him to continue to do problem solving. Therefore, he likes easier problems.

In response to survey question (33), close to $90\%$ of the graduate students agreed that two introductory level 
problems, both of which involve conservation of energy, can be solved using similar methods 
whereas only approximately $55\%$ of them agreed that both problems can be solved using similar methods
when solving graduate level conservation of energy problems (see Figure 1). Individual discussions with a subset of graduate students suggest that they
felt that since air-resistance and friction were involved, they may have to use different methods to solve the problems.
In particular, they noted that they often use different methods involving Lagrangians and Hamiltonians 
to solve complicated problems
in graduate level courses and they were not sure if the same technique will be useful in problems involving friction and air-resistance. 
In response to survey question (33), all of the physics faculty noted that both problems can be solved using similar methods (see Table 1). 
The responses of many graduate students to question (33) points to the fact that graduate
students who are taking graduate level physics courses are immersed in learning complicated mathematical techniques and
they are evaluating their survey responses in light of their experiences with mathematical tools. When a physics faculty member, who had taught
several ``core" graduate courses routinely, was shown the responses of graduate students to question (33), he commented that he has observed 
that, sometimes the graduate students are so focused on mathematical
manipulations in the graduate-level courses, they tend to use unnecessarily complicated techniques even when they are asked questions
which can be solved using introductory level techniques, e.g., for problems related to Gauss's law or Ampere's law.

\subsection{Comparison of graduate students' survey responses with those of other groups}

Next, we compare graduate students' responses on selected questions on the AAPS survey
with those of the physics faculty and introductory physics and astronomy students.

\subsubsection{Graduate Students are still developing expertise in graduate level problem solving}

In the previous section, we discussed that the average responses of graduate students to graduate-level problem solving was less expert-like
than their responses to introductory level problem solving.
Comparison of average graduate students' responses for graduate level problem solving with those of physics faculty also suggests
that graduate students are still developing expertise in problem solving at the graduate level. For example,
Figure 2 shows that, in response to question (6), all of the physics faculty noted that while solving physics problems they could often tell when their work and/or answer is wrong even
without external resources but only approximately $50\%$ of the graduate students could do so while solving graduate level problems
and approximately $80\%$ of the graduate students could do so for introductory level problem solving. Moreover, the survey response of the graduate students to
this question for graduate level problems is similar to that of the introductory physics students for introductory
level problems. Such similarity suggests that while graduate students may be experts in solving introductory problems, 
they are still developing expertise in solving graduate level problems. 

Figure 3 shows that, in response to question (11) about whether equations need not be intuitive in order to be used and whether
they routinely use equations even if they are non-intuitive, graduate students' responses while solving introductory
physics problems were similar to those of faculty and approximately $75\%$ disagreed with the statements (favorable response). 
However, when answering graduate level problems,
only slightly more than $50\%$ of the graduate students noted that equations must be understood in an intuitive sense before being used.
Individual discussions suggest that the graduate
students felt that sometimes the equations encountered in the graduate courses are too abstract and they do not have sufficient time
to make sense of them and ensure that they have built an intuition about them. 
The following sample responses from some graduate students reflect their sentiments:
\begin{itemize}
\item {\it``...you just cannot understand everything. So it's ok to deal with the homework
first. But I really feel bad when I do plug and chuck [sic]."} 
\item {\it``I am often still presented with
equations to calculate something without enough motivation to understand the process, even at the graduate level, and being able to
use the equation and accept that you'll understand it later is often
necessary.  For students' first course in physics, this is more the rule than the exception at some level..."}
\item {\it``I remember physics via the equations, so I try my best to always
understand the meaning. But if I can't, I fall back on ``this is the equation, use it"."}
\item {\it ``As an introductory student I had the point of view that the equations are  right so my intuition must be wrong.  
I used equations to get the answer whether it made sense at first or not, but I trained my intuition with every such result. I had more faith 
in the physics  that is taught to me than the physics intuition I acquired just by observation. As a graduate student, one is already used to the  
unintuitive results being the correct one, they have by then become intuitive."}
\end{itemize}
The last graduate student quoted above expresses an interesting view that by the time one becomes a graduate student in physics, one may have
learned to accept non-intuitive results and such results start appearing intuitive. The responses of the introductory physics 
students suggest that they are even more likely than graduate students to use equations to calculate answers even if they are 
non-intuitive (see Figure 3).
This finding is consistent with the prior results that suggest that many introductory students view problem solving in physics as an exercise in finding
the relevant equations rather than focusing on why a particular physics principle may be applicable and building an intuition about a certain type of
physics problems.~\cite{mpex,class1}

In response to question (25) about whether individuals make sure they learn from their mistakes and do not make the same mistakes 
again, all but one physics faculty agreed with the statement (favorable) and one was neutral. 
On the other hand, only slightly more than $60\%$ and $70\%$ of the graduate students agreed with the statement when pertaining to solving graduate level problems and introductory level problems, respectively. Graduate students's net favorable scores for question (25) in Table 1 suggests
that many graduate students had unfavorable response to this question especially for graduate-level problem solving.

The response of introductory physics students to question (25) was comparable to that of graduate students for introductory physics 
problem solving but it is less expert-like than the physics faculty.
One introductory student said he did not review errors on the midterm exam as much as he would on homework, partly 
because the homework problems may show up on a future test but partly because he didn't like staring at his bad exam grade.  
The reluctance to reflect upon tests is consistent with our earlier findings for an upper-level undergraduate quantum mechanics course which
demonstrated that many students did not reflect automatically on their mistakes in the midterm exams for similar reasons.~\cite{quantum}

\subsubsection{Unexpected Trends require Careful Analysis}

The average responses to some survey questions for various groups are counter-intuitive and require careful analysis.
Figure 4 shows that on question (5) of the survey,  while no faculty agreed with the statement (no unfavorable response) that problem solving in physics basically means matching problems with the correct equations and then substituting values to get a number, more than $30\%$ of the graduate students agreed with the statement (unfavorable) in the context of introductory level problem solving and approximately $20\%$ agreed with the statement for graduate level problem solving.
However, Figure 4 also shows the counter-intuitive trend that the average responses of introductory physics students to question (5)
were indistinguishable from those of the graduate students for introductory physics problem solving. Individual discussions and written
explanations suggest that the reasoning of many graduate students differs from the reasoning of a typical introductory physics student although the
average responses of these groups are superficially the same. 
In particular, for introductory level problem solving, many graduate students felt so comfortable with the applications of
basic principles that not much explicit thought was involved in solving introductory level problems. 
For example, in response to question (5), one graduate student noted
\begin{itemize}
\item {\it ``Well for introductory physics this is true. But, in more advanced problems you kind of have to setup the equations."}
\end{itemize}
On the other hand, prior research suggests that many introductory physics students think that physics is a collection of disconnected facts and 
formulas and use a ``plug and chug" approach to problem solving without thinking if a principle is applicable in a particular 
context.~\cite{mpex,class1} 

Some graduate students reflected explicitly on their introductory physics experiences and compared it to the graduate level experiences
in problem solving. The following is a reflective response of one graduate student 
\begin{itemize}
\item {\it``you can get an expression from two others without understanding how or why.  As an introductory student I probably did  
this more because the expressions were simpler and easier to  manipulate without a $100\%$ understanding. My motives were also more to  
get the work done than to learn every detail."}
\end{itemize}

Figure 5 shows that, in response to question (14) regarding whether they always explicitly think about concepts that underlie the problems when solving physics problems,
close to $90\%$ of the graduate students agreed (favorable) that they do so both in the context of introductory and graduate level problem solving. However, only approximately
$65\%$ and $55\%$ of the physics faculty and introductory physics students agreed, respectively. The 
trend in Figure 5 going from the introductory students to faculty is not consistent with expectations at first, but
individual discussions suggest that some faculty do not always explicitly think about the concepts that underlie the problem because the concepts
have become obvious to them due to their vast experience. They are able to invoke the relevant physics principles, e.g., conservation of mechanical
energy or conservation of momentum, automatically when solving an introductory problem without making a conscious effort. 
In fact, question (14) is one of those rare questions on the survey for which the faculty responses were different for 
introductory and graduate-level problem solving (in particular, more than $90\%$ noted that they explicitly think about
concepts that underlie the problems for graduate-level problems).
In contrast, prior
research suggests that introductory physics students often do not explicitly think about the relevant concepts because they often consider physics as
consisting of disconnected facts and formulas and associate physics problem solving as a task requiring hunting for the relevant formulas
without performing a conceptual analysis and planning of the problem solution.~\cite{mpex,class1} Thus, the reasoning behind the less favorable
responses of faculty to question (14) is generally very different from the reasonings behind the introductory physics students' responses.

\subsubsection{Introductory physics students are not on par with physics graduate students and faculty}

Survey responses to some questions suggest that the introductory physics students' attitudes and approaches to solving
introductory physics problems are not as expert-like as physics graduate students and faculty. For example, survey responses suggest that
manipulation of symbols rather than numbers increases the difficulty of a problem for many introductory physics students. Question (30) asked whether symbolic
problems were more difficult than identical problems with numerical answers and question (31) asked if individuals preferred to
solve a problem with a numerical answer symbolically first and only plug in the numbers at the very end. Figures (6) and (7) show that the 
responses of graduate students for both introductory and graduate level problem solving are comparable to physics faculty but introductory
students' responses are very different. In response to question (30), only approximately $35\%$ of the introductory physics students disagreed with the statement (favorable response) that it is more difficult to solve a problem symbolically, and in response to question (31), only
$45\%$ agreed with the statement (favorable response) that they prefer to solve the problem symbolically first and only plug in the numbers at the very end. 

Individual discussions with some introductory physics students suggest that they
have difficulty keeping track of the variables they are solving for if they have several symbols floating around, which motivates them to substitute
numbers at the beginning of the solutions.~\cite{torigoe} 
Some introductory students noted that they did not like carrying expressions involving symbols from one equation
to another because they were afraid that they will make mistakes in simplifying the expressions.

\subsubsection{Graduate student attitudes and approaches about introductory physics problem solving is not always as expert-like as physics faculty}

The responses to some survey questions suggest that the attitudes and approaches of graduate students regarding solving introductory level problems 
is not as expert-like as physics faculty. For example,
Figure 8 shows that, in response to question (2) about whether they often make approximations about the physical world when solving physics problems, all faculty noted that they
do so. However, only approximately $75\%$ and $65\%$ of graduate students noted they do so for graduate level problem solving and introductory
level problem solving, respectively. Individual discussions and written explanations suggest that the graduate students have different
views about making approximations about the physical world as illustrated by the sample comments below:
\begin{itemize}
\item {\it ``I don't connect the physics problems to real world very much."}
\item {\it ``it's stat mech, in which I do whatever I have to} [including approximations], 
{\it to make the answer come out (and usually that is correct)."} 
\item {\it ``Solving physics problems as an introductory physics student I was  
perhaps more prone to this, thinking about how a block would slide  down an incline. As I became more familiar with the extent of 
``non-physical" approximations we made such as a frictionless world, I learned to separate problem solving space and real life space. I find  
that this is one aspect of physics problem solving that is harder for introductory level courses than graduate courses, the problems we  
solve} [in introductory physics] {\it are farther away from the physical world than graduate level  
problems. It keeps the math manageable and the physics concepts manageable but it makes them less intuitive."}
\item {\it ``Many introductory-level problems are well defined and ideal, which doesn't require approx."}  
\end{itemize}

In contrast to the last graduate student's comment, individual discussions with some faculty suggest that they considered idealization of the 
problem in introductory and graduate level physics, e.g., framing problems without friction or air resistance, considering spherical cows or 
point masses, the infinite square well or Hydrogen atom with only the Coulomb force, etc. as making 
approximations about the physical world and they felt that such approximations were helpful for getting
an analytical answer and for building intuition about physical phenomena.
It is possible that approximately $25\%$ of the graduate students who noted that they don't make approximations while solving introductory
physics problems have not carefully thought about the role of approximations about the physical world in introductory physics problem solving.

Figure 9 shows that, in response to question (12) regarding whether physics involves many equations each of which applies primarily to a specific 
situation, all but one physics faculty members disagreed with the statement (favorable) but less than $70\%$ of the graduate students disagreed with it when solving 
graduate level problems. The percentage of introductory physics students who disagreed with the statement was slightly more than $35\%$ and slightly 
more than $40\%$ for the astronomy students. These responses are commensurate with the expertise of each group and points to the fact that
experts are more likely to discern the coherence of the knowledge in physics and appreciate how very few laws of physics are applicable in diverse situations and
can explain different physical phenomena.

\subsubsection{Some questions do not have a clear-cut expert-like response}

If faculty are taken as experts, their responses to the survey questions indicate that there are 
some questions whose answers may not necessarily represent clear-cut favorable/unfavorable traits without disagreement from several faculty. 
For example, for question (24), there are differences in faculty responses for introductory-level vs graduate-level problem solving.
For example, less than $70\%$ of the faculty (and an even smaller percentage of graduate students)
noted that they liked to think through a difficult physics problem with a peer when solving introductory physics problem but
$83\%$ of the faculty liked to work with a peer for difficult graduate level problems. 
Individual discussions with some of the faculty members suggests that whether one continues to persevere
individually or works with a peer to solve a challenging introductory or graduate-level problem depends on an individual's personality.
The graduate student reasonings for wanting to work with a peer or not were varied as illustrated by the following examples:
\begin{itemize}
\item {\it ``This is not true} (usefulness of talking to peers) {\it for introductory-level problems 
because, typically these types of problems are quite direct.  Thinking about them for a little
while always produces some result.  For graduate-level problems, it is almost essential to work with others because of the complex line of
thought it takes to solve some problems."} 
\item {\it ``Bouncing ideas with someone gives me sort of a 
chance to see the problem from the outside. You somehow see another point of attack. If you are stuck on a problem, usually the reason is that 
the approach is a dead end or too complex. Having someone to talk to forces you to think with a different perspective."}
\item {\it ``I would like to think out by myself."}
\item {\it ``As an introductory student it} (working with peers) {\it can make things more complicated, I  
would rather ask the TA.  Other students have the same misconceptions as I do so they aren't a good source. As a graduate student, I saw  
that everyone was benefiting from collaboration. I know I would too. I just don't like to do anything with a peer but that's purely a  
social issue, I believe it is useful to work with peers."} 
\end{itemize}
Thus, while many graduate students agree that talking to peers is helpful (at least for 
challenging problems) some of them are inherently more averse to discussions with peers than others.

Figure 10 shows that in response to question (3), regarding whether being able to handle the mathematics is the most important part of the process in
solving a physics problem, less than $60\%$ of the physics faculty disagreed with the statement (favorable response) and approximately $35\%$ were
neutral. Amongst graduate students, less than $40\%$ of the students disagreed with the statement (favorable response) both for problem solving
at the introductory and graduate levels. Roughly $50\%$ of the graduate students agreed (unfavorable response) that mathematics is the most important
part of the process in problem solving
at the graduate level and more than $40\%$ agreed (unfavorable response) with it for introductory level problem solving. 
Individual discussions with the graduate students suggest that, in response to question (3), some students felt that facility with high level 
mathematics is the most important skill for excelling in their graduate courses. Some graduate students 
felt that basic mathematics is very important for doing well in introductory problem solving as well, whereas others did not think 
mathematics was as important, especially for the algebra-based courses.
The following are examples of responses from the graduate students that convey the sentiments:
\begin{itemize}
\item {\it ``if I was teaching a class of med. students, the concepts are certainly most important...  
However, a class of engineers really need to know how to get the right answer so the things
that they build, function as they are supposed to.  I would say in this case that math and concepts are equally important. For graduate-level 
problems, I believe mathematics becomes more essential over introductory-level problems."}
\item {\it ``From my point of view, the introductory physics concepts are very easy to 
understand. It's in the details of problem solving that you could get 
stuck like encountering a difficult integration for example or some tricky algebra."}
\item {\it ``In introductory physics, this was not at all the case, the math was  
easy enough, I needed little more than high school calculus, so it was  
getting the physics down and understanding the language of physics,  
the new jargon, new concepts etc.  Once the concepts became familiar  
enough and I moved on to graduate school, math became my biggest  
problem.  From vector calculus to advanced linear algebra to special  
functions to group theory, the math is often harder.  I find it a lot  
easier to think about physics and the universe conceptually (now that  
I am armed with such intuition and interest) but trying to actually  
`solve' a physics problem comes down to the math, which I find hard."}   
\end{itemize}
Individual discussions with some physics faculty about question (3) suggests that they believed that conceptual knowledge in physics
was the central aspect of physics problem solving in both introductory and graduate level problem solving.
But some faculty who were neutral in response to question (3) emphasized that the students may not excel in physics 
without a good grasp of mathematics even though concepts are central to learning physics. 
The views of more graduate students (compared to faculty) about 
mathematics being the most important aspect of physics problem solving may stem from the fact that graduate students have recently taken
graduate and undergraduate level courses in which their grades often depend not as much on their conceptual knowledge but on
their mathematical facility. However, question (3) is one of the survey questions for which
there isn't a strong agreement on favorable response amongst the faculty either (see Figure 10).

Similarly, in response to question (16), only $75\%$ of faculty for introductory level and a somewhat higher percentage for graduate level
(and a much smaller percentage of graduate students in each case) noted that, while answering 
conceptual physics questions, they use the physics principles they usually think about when solving quantitative problems rather than 
mostly using their ``gut" feeling. Discussions elucidated that the faculty members' use of their ``gut" feeling to answer conceptual
questions (rather than explicitly invoking physics principles) was often due to the fact that they had developed
good intuition about the problems based upon their vast experience.
Thus, they did not need to explicitly think about the physical principles involved.  

Incidentally, in response to question (16), $50\%$ of the introductory physics
students claimed that they use their ``gut" feeling to answer conceptual questions rather than invoking physics
principles. Our earlier research and those of others suggest that introductory students often view conceptual
questions as guessing tasks and use their ``gut" feeling rather than explicitly considering how the physical
principles apply in those situations.~\cite{mpex,eric,our} 
One interviewed introductory student stated that he would not consider principles when answering a 
conceptual question because over-analyzing the problem is more likely to make his answer wrong.  
When Mazur from Harvard University gave the Force Concept Inventory Conceptual standardized test~\cite{fci} to his introductory students, 
a student asked if he should do it the way he really thinks about it or the way he has been taught to think about it in the class.~\cite{eric} It appears that students sometimes hold two views simultaneously, where one is based upon their gut feeling and another is based upon what
they learned in the physics class, and these views coexist and are difficult to merge.

\subsubsection{Why don't faculty and graduate students reflect after solving a problem?}

Problem solving is often a missed learning opportunity because, in order to learn from problem solving, one must reflect upon the problem 
solution.~\cite{black1,black2,edit1,edit2,andy2,quantum}
For example, one must ask questions such as ``what did I learn from solving this problem?", ``why did the use of one principle work and not the other one?" or
``how will I know that the same principle should be applicable when I see another problem with a different physical situation?".
Unfortunately, the survey results suggest a general lack of reflection by individuals in each group after solving problems. Figure 11 shows that in response to 
question (20), only approximately $55\%$ of the graduate students (for both introductory and advanced level problems) 
noted that, after they solve homework problems, they take the time to reflect and learn from the solution. The percentage of faculty
members who noted that they take the time to reflect is close to $75\%$ (for both introductory and graduate-level problem solving)
which appears to be lower than expected. 
Individual discussions suggest that physics faculty felt that they monitor their thought processes while solving the problems.
Therefore, reflection at the end of problem solving is not required. In contrast, while solving graduate level homework problems,
some graduate students pointed to the lack of time for why they do not take the time to reflect after solving problems. 
Following are some explanations from graduate students for their responses:
\begin{itemize}
\item  {\it ``If I have enough time, then I would like to reflect and learn from the problem solution after I
struggle with it for a long time and then finally solve it successfully."} 
\item {\it``If the solution or the problem is
interesting, then I would take time to reflect and learn from it. This usually happens in more challenging problems."} 
\item {\it ``To be honest, I didn't do this when I was in college. But now I realized it's helpful."}
\end{itemize}
Only approximately $25\%$ of introductory physics students noted that they reflect and learn from problem solutions. Since reflection is so important 
for learning and building a robust knowledge structure, these findings suggest that instructors should consider giving students explicit incentive to reflect after they solve physics 
problems.~\cite{black1,black2,edit1,edit2,andy2,quantum}

\section{Summary and Conclusion}

We developed and validated the ``Attitudes and Approaches to Problem Solving" (AAPS) survey based upon an earlier Attitudes towards problem solving survey by Cummings et. al.
The survey was administered  to physics graduate students, who answered the survey questions
about problem solving in their graduate courses and in introductory physics. Their survey responses were
compared with those of introductory students in physics and astronomy courses and physics faculty.
We discussed the responses individually with some students and faculty and obtained written explanations from some graduate students on selected questions.

There were major differences on some measures in graduate students' responses about problem solving in the graduate courses
compared to problem solving in introductory physics. In general, graduate students' responses about problem solving in the graduate courses were less
favorable (less expert-like) than their responses about solving introductory physics problems. 
For example, graduate students were more likely to feel stuck unless they got help while solving graduate level problems than
on introductory level problems. Similarly, for problem solving in their graduate level courses,
fewer graduate students could tell when their work and/or answer was wrong without talking to someone
else but many more could tell when their solution was not correct when solving introductory physics problems.
Also, more graduate students noted that they routinely use equations even if they are 
non-intuitive while solving graduate level problems than while solving introductory physics problems.
In addition, fewer graduate students noted that they enjoy solving challenging graduate
level physics problems than solving challenging introductory physics problems (perhaps because introductory physics problems
are still easier for them). 

Comparison of graduate students' responses with faculty responses suggests that, on several measures, graduate students' responses to AAPS survey
are less expert-like than faculty responses. For example, unlike the graduate students, all physics faculty noted that they enjoy solving 
challenging physics problems. The less favorable response of graduate students while solving graduate level problems is partly due to
the fact that the graduate students are not yet experts, especially in their own graduate course content. Due to lower expertise in
solving graduate level problems, graduate students are more likely to feel stuck unless they get help,
not know whether their solution is right or wrong, use equations that are not intuitive and not enjoy solving challenging
graduate level problems on which their grade depends and for which they have a limited time to solve. 

We find that, on some survey questions, graduate students' and faculty responses to the survey questions must be interpreted carefully. 
For example, only two thirds of the faculty noted that they always think about the concepts that underlie the problem explicitly
while solving introductory-level problems,
which is lower than the fraction of graduate students who noted that they do so while solving both introductory level and
graduate level problems. Individual discussions with faculty members suggests that they felt that, after years of teaching experience, the concepts
that underlie many of the introductory physics problems have become ``automatic" for them and they do not need to explicitly think about them. The fact that
in contrast to most faculty, many graduate students always think explicitly about the concepts that underlie the problems both while solving introductory
and graduate level problems, suggests that the graduate students have not developed the same level of expertise and automaticity in solving
introductory level problems as physics faculty have.
In fact, question (14) related to this issue is one of those rare questions on the survey for which the faculty responses were significantly
different for introductory and graduate-level problem solving (in particular, more than $90\%$ of the faculty noted that they explicitly think about
concepts that underlie the graduate-level problems while solving them).

Comparison of graduate students' AAPS responses with introductory physics students' responses suggests that, on some measures, graduate students 
have more favorable attitudes and approaches to solving introductory physics problems due to their higher level of expertise than
the introductory students. However, on other questions, the responses must be interpreted carefully in light of the explanations provided
by the graduate students. For example, in response to whether the problem solving in physics is essentially ``plug and chug", the response
of the graduate students while solving introductory physics problems and those of introductory physics students is indistinguishable.
But discussions and written explanations of graduate students suggest that they have developed sufficient expertise in introductory physics 
so that solving such problems does not require much explicit thought and they can often immediately tell which
principle of physics is applicable in a particular situation. 
On the other hand, prior research suggests that many introductory physics students jump into implementation of problem solution
and immediately look for the formulas without performing a conceptual analysis and planning of the problem solution.~\cite{reif1,reif2,reif3} 

Also, due to their higher level of expertise, graduate students find introductory physics equations more intuitive and are better able to
discern the applicability of a physics
principle epitomized in the form of a mathematical equation to diverse situations than the introductory students. In solving both introductory and 
graduate level problems, the fraction of graduate students who noted that they reflect and learn from the problem solution after
solving a problem is larger than the fraction of introductory physics students who noted doing so in their courses. While we may desire
an even higher percentage of graduate students to reflect and learn from their problem solving, written explanations suggest that, in the graduate
courses, some students felt they did not have the time to reflect. Also,
they often did not reflect on the exam solutions even after they received the solutions because they did not expect those
problems to show up again on another exam. Some graduate students explained that the reason they do not reflect after solving
an introductory physics problem is that the solutions to those problems are obvious to them and do not require reflection.

There was a large difference between the introductory physics students' and graduate students' responses in their facility to manipulate 
symbols (vs. numbers) with introductory physics students finding it more difficult to solve problems given in symbolic form. In problems where 
numbers were provided, many introductory students noted that they prefer to plug numbers at the beginning rather than waiting till the end to
do so.  One suggested strategy to help introductory physics students feel more confident about using symbols is to ask them to underline the variable they 
are solving for so as to keep it from getting mixed up with the other variables.~\cite{reifbook}
Developing mathematical facility can also help students develop the confidence to solve the problems symbolically first before substituting values. In addition, instructors can emphasize
why it is useful to keep the symbols till the end, including the fact that it can allow them to check the correctness of the solution, e.g., by
checking the dimension, and it can also allow them to check the limiting cases which is important for developing confidence in one's solution.~\cite{reifbook}

In general, the more favorable responses of graduate students on the AAPS survey towards attitudes and approaches to introductory problem solving compared to 
those of the introductory physics students and less favorable responses compared to the faculty imply that graduate students have a higher level 
of expertise in introductory physics  but less expertise than physics faculty. Moreover,
graduate students' responses to graduate level problem solving in many instances are comparable to introductory students' responses to
introductory level problem solving, implying that the graduate students are still developing expertise in their own graduate level courses
just like introductory students are still developing expertise in introductory physics.

As noted earlier, the survey results also suggest that many graduate students are more likely to enjoy solving difficult introductory physics problems
than graduate level problems, they are more likely to feel stuck while solving graduate level problems, less likely to 
find graduate level equations intuitive (but they still use them freely to solve problems), 
less likely to predict whether their problem solution is correct and to not give a high priority to reflecting and learning after solving a problem.
While one can rationalize these less expert-like responses of graduate students to graduate level problem solving by claiming that these are 
reflections of the fact that
they are not ``experts" in graduate level courses, they force us to think about whether we are achieving the goals of the graduate
courses and giving graduate students an opportunity to learn effective approaches and attitudes to problem solving. Graduate instructors
should consider whether assessment in those courses should include both quantitative and conceptual questions to motivate students to reflect on 
problem solving and give explicit incentive for reflection and for development of intuition about the equations underlying the problems.

\section{Acknowledgments}

We thank F. Reif, R. Glaser, R. P. Devaty and J. Levy for very useful discussions.
We thank Dr. Clement Stone for help with statistical analysis.
We are grateful to all the faculty who administered the survey to their class and took the survey for their help.

\bibliographystyle{aipproc}

\pagebreak

\begin{center}
{\bf Appendix: AAPS Survey Questionnaire and favorable (expert) responses}
\end{center}

\noindent
To what extent do you agree with each of the following statements when you solve physics 
problems? Answer with a single letter as follows:\\
\noindent
                   A)  Strongly Agree\\
                   B)  Agree Somewhat\\
                   C)  Neutral or Don't Know\\
                   D)  Disagree Somewhat\\
                   E)  Strongly Disagree\\

\noindent
1.  If I'm not sure about the right way to start a problem, I'm stuck unless I go see the teacher/TA 
or someone else for help. (D/E)

\noindent
2.  When solving physics problems, I often make approximations about the physical world. (A/B)

\noindent
3.   In solving problems in physics, being able to handle the mathematics is the most important 
part of the process. (D/E)

\noindent
4.  In solving problems in physics, I always identify the physics principles involved in the problem 
first before looking for corresponding equations. (A/B)

\noindent
5.   "Problem solving" in physics basically means matching problems with the correct equations 
and then substituting values to get a number. (D/E)

\noindent
6.   In solving problems in physics, I can often tell when my work and/or answer is wrong, even 
without looking at the answer in the back of the book or talking to someone else about it. (A/B)

\noindent
7.  To be able to use an equation to solve a problem (particularly in a problem that I haven't seen 
before), I think about what each term in the equation represents and how it matches the 
problem situation. (A/B)

\noindent
8.   There is usually only one correct way to solve a given problem in physics. (D/E)

\noindent
9.   I use a similar approach to solving all problems involving conservation of linear momentum 
even if the physical situations given in the problems are very different. (A/B)

\noindent
10. If I am not sure about the correct approach to solving a problem, I will reflect upon physics 
principles that may apply and see if they yield a reasonable solution. (A/B)

\noindent
11. Equations are not things that one needs to understand in an intuitive sense; I routinely use 
equations to calculate numerical answers even if they are non-intuitive. (D/E)

\noindent
12. Physics involves many equations each of which applies primarily to a specific situation. (D/E)

\noindent
13. If I used two different approaches to solve a physics problem and they gave different answers, 
I would spend considerable time thinking about which approach is more reasonable. (A/B)

\noindent
14. When I solve physics problems, I always explicitly think about the concepts that underlie the 
problem. (A/B)

\noindent
15. When solving physics problems, I often find it useful to first draw a picture or a diagram of the 
situations described in the problems. (A/B)

\noindent
16.  When answering conceptual physics questions, I mostly use my ``gut" feeling rather than 
using the physics principles I usually think about when solving quantitative problems. (D/E)

\noindent
17.  I am equally likely to draw pictures and/or diagrams when answering a multiple-choice 
question or a corresponding free-response (essay) question. (A/B)

\noindent
18.  I usually draw pictures and/or diagrams even if there is no partial credit for drawing them. (A/B)

\noindent
19.  I am equally likely to do scratch work when answering a multiple-choice question or a 
corresponding free-response (essay) question. (A/B)

\noindent
20.  After I solve each physics homework problem, I take the time to reflect and learn from the 
problem solution. (A/B)

\noindent
21.  After I have solved several physics problems in which the same principle is applied in 
different contexts, I should be able to apply the same principle in other situations. (A/B)

\noindent
22.  If I obtain an answer to a physics problem that does not seem reasonable, I spend considerable 
time thinking about what may be wrong with the problem solution. (A/B)

\noindent
23.  If I cannot solve a physics problem in 10 minutes, I give up on that problem. (D/E)

\noindent
24.  When I have difficulty solving a physics homework problem, I like to think through the 
problem with a peer. (A/B)

\noindent
25.  When I do not get a question correct on a test or homework, I always make sure I learn from 
my mistakes and do not make the same mistakes again. (A/B)

\noindent
26.  It is more useful for me to solve a few difficult problems using a systematic approach and 
learn from them rather than solving many similar easy problems one after another. (A/B)

\noindent
27.  I enjoy solving physics problems even though it can be challenging at times. (A/B)

\noindent
28.  I try different approaches if one approach does not work. (A/B)

\noindent
29.  If I realize that my answer to a physics problem is not reasonable, I trace back my solution to 
see where I went wrong. (A/B)

\noindent
30.  It is much more difficult to solve a physics problem with symbols than solving an identical 
problem with a numerical answer. (D/E)

\noindent
31.  While solving a physics problem with a numerical answer, I prefer to solve the problem 
symbolically first and only plug in the numbers at the very end. (A/B)

\noindent
32.  Suppose you are given two problems.  One problem is about a block sliding down an inclined 
plane with no friction present. The other problem is about a person swinging on a rope.  Air 
resistance is negligible. You are told that both problems can be solved using the concept of 
conservation of mechanical energy of the system.  Which one of the following statements do 
you MOST agree with?  (Choose only one answer.) (A/B)

\noindent
A)  The two problems can be solved using very similar methods.\\
B)  The two problems can be solved using somewhat similar methods.\\
C)  The two problems must be solved using somewhat different methods.\\
D)  The two problems must be solved using very different methods.\\
E)  There is not enough information given to know how the problems will be solved. 

\noindent
33. Suppose you are given two problems.  One problem is about a block sliding down an inclined 
plane. There is friction between the block and the incline.  The other problem is about a 
person swinging on a rope. There is air resistance between the person and air molecules. You 
are told that both problems can be solved using the concept of conservation of total (not just 
mechanical) energy. Which one of the following statements do you MOST agree with?  
(Choose only one answer.) A/B

\noindent
A)  The two problems can be solved using very similar methods.\\
B)  The two problems can be solved using somewhat similar methods.\\
C)  The two problems must be solved using somewhat different methods.\\
D)  The two problems must be solved using very different methods.\\
E)  There is not enough information given to know how the problems will be solved.

\pagebreak

\begin{table}[h]
\centering
\begin{tabular}[t]{|c|c|c|c|c|c|c|c|}
\hline
Problem number& 1& 2& 3& 4& 5& 6& 7 \\[0.5 ex]
\hline
Faculty-Intro& 0.83& 1.00& 0.50& 0.92& 0.92& 1.00& 0.83 \\[0.5 ex]
\hline
Faculty-Grad& 1.00& 0.92& 0.36& 1.00& 1.00& 1.00& 1.00 \\[0.5 ex]
\hline
Graduate Students-Intro& 0.71& 0.42& -0.04& 0.83& 0.17& 0.75& 0.83 \\[0.5 ex]
\hline
Graduate Students-Self& 0.40& 0.63& -0.13& 0.75& 0.63& 0.25& 0.88 \\[0.5 ex]
\hline
Astronomy Students& 0.45& 0.48& -0.16& 0.58& 0.13& 0.71& 0.84 \\[0.5 ex]
\hline
All Introductory Students& 0.14& 0.19& 0.15 & 0.41& 0.16& 0.24& 0.61 \\[0.5 ex]
\hline
\hline
Problem number& 8& 9& 10& 11& 12& 13& 14\\[0.5 ex]
\hline
Faculty-Intro& 0.92& 0.58& 0.92& 0.67& 0.83& 1.00& 0.50\\[0.5 ex]
\hline
Faculty-Grad& 1.00& 0.92& 1.00& 0.75& 1.00& 0.92& 0.92\\[0.5 ex]
\hline
Graduate Students-Intro& 0.83& 0.46& 0.88& 0.67& 0.54& 0.88& 0.88\\[0.5 ex]
\hline
Graduate Students-Self& 1.00& 0.31& 0.69& 0.33& 0.44& 0.94& 0.81\\[0.5 ex]
\hline
Astronomy Students& 0.77& 0.35& 0.94& 0.23& 0.10& 0.74& 0.77\\[0.5 ex]
\hline
All Introductory Students& 0.67& 0.24& 0.58& -0.03& -0.06& 0.56& 0.32\\[0.5 ex]
\hline
\hline
Problem number& 15& 16& 17& 18& 19& 20& 21\\[0.5 ex]
\hline
Faculty-Intro& 1.00& 0.67& 1.00& 0.92& 1.00& 0.75& 1.00\\[0.5 ex]
\hline
Faculty-Grad& 0.92& 0.75& 0.92& 0.92& 1.00& 0.75& 1.00\\[0.5 ex]
\hline
Graduate Students-Intro& 0.96& 0.50& 0.79& 0.96& 0.88& 0.38& 0.92\\[0.5 ex]
\hline
Graduate Students-Self& 0.94& 0.31& 0.50& 0.88& 0.56& 0.25& 0.94\\[0.5 ex]
\hline
Astronomy Students& 0.29& 0.52& 0.06& 0.19& 0.84& 0.32& 0.90\\[0.5 ex]
\hline
All Introductory Students& 0.74& 0.23& 0.55& 0.69& 0.77& -0.19& 0.71\\[0.5 ex]
\hline
\hline
Problem number& 22& 23& 24& 25& 26& 27& 28\\[0.5 ex]
\hline
Faculty-Intro& 1.00& 0.92& 0.42& 0.92& 1.00& 0.92& 1.00\\[0.5 ex]
\hline
Faculty-Grad& 1.00& 1.00& 0.67& 1.00& 0.92& 1.00& 1.00\\[0.5 ex]
\hline
Graduate Students-Intro& 1.00& 1.00& 0.21& 0.54& 0.71& 0.67& 0.96\\[0.5 ex]
\hline
Graduate Students-Self& 1.00& 0.75& 0.19& 0.38& 0.50& 0.63& 0.88\\[0.5 ex]
\hline
Astronomy Students& 0.77& 0.74& 0.06& 0.68& 0.55& 0.74& 0.87\\[0.5 ex]
\hline
All Introductory Students& 0.52& 0.40& 0.43& 0.56& 0.37& 0.03& 0.75\\[0.5 ex]
\hline
\hline
Problem number& 29& 30& 31& 32& 33&Avg. &   \\[0.5 ex]
\hline
Faculty-Intro&  1.00& 0.93& 1.00 & 1.00 & 1.00&0.88 &  \\[0.5 ex]
\hline
Faculty-Grad&  1.00& 1.00& 1.00 & 1.00 & 1.00&0.92 &  \\[0.5 ex]
\hline
Graduate students-Intro& 1.00& 0.92& 0.92& 1.00& 0.83&0.73 & \\[0.5 ex]
\hline
Graduate students-Self& 1.00& 0.69& 1.00& 0.88& 0.19&0.62 & \\[0.5 ex]
\hline
Astronomy Students& 0.68& -0.19& -0.13& 0.68& 0.50&0.49 & \\[0.5 ex]
\hline
All Introductory Students& 0.74& -0.04& 0.08& 0.70& 0.46&0.33 & \\[0.5 ex]
\hline
\hline
\end{tabular}
\vspace{0.1in}
\caption{Average scores for each group of students and faculty on each of the individual questions and averaged
over all survey questions (see last entry). To calculate the average score for a question, a +1 is assigned to each favorable 
response, a -1 is assigned to each unfavorable response, and a 0 is assigned to neutral responses. One then averages these values for everybody in 
a particular group (e.g., faculty-Intro) to obtain an average score for that group. ``Intro" and ``Self" with Graduate students
implies problem solving in ``introductory physics" and ``graduate-level physics courses" respectively.
}
\label{junk2}
\end{table}

\pagebreak

\begin{table}[h]
\centering
\begin{tabular}[t]{|c|c|c|c|c|c|c|}
\hline
Cohen's d&Intro physics&Astronomy&Grad-intro&Grad-self&Faculty-intro&Faculty-grad\\[0.5 ex]
\hline
Intro physics (541)&&0.62&1.60&1.30&2.18&2.36\\[0.5 ex]
\hline
Astronomy (31)& & &1.42&0.90&2.19&2.43\\[0.5 ex]
\hline
Grad-intro (42)&&&&0.46&1.19&1.53\\[0.5 ex]
\hline
Grad-self (34)&&&&&1.43&1.71\\[0.5 ex]
\hline
Faculty-intro (12)&&&&&&0.48\\[0.5 ex]
\hline
Faculty-grad (12)&&&&&&\\[0.5 ex]
\hline
\end{tabular}
\vspace{0.1in}
\caption{
Effect sizes between two groups.  The number of people in each group is given in the parenthesis.
All effect sizes are positive because they are always taken such that $\mu_2<\mu_1$ in the calculations (i.e. subtracting the higher mean from the lower mean; see Table 1). 
ANOVA using Pairwise t-test shows that the differences between all the groups is significant except that between faculty members for introductory-level and
graduate-level problem solving (p=0.269 for that case).
}
\label{junk4}
\end{table}

\begin{figure}[h!]
  \includegraphics[height=.3\textheight]{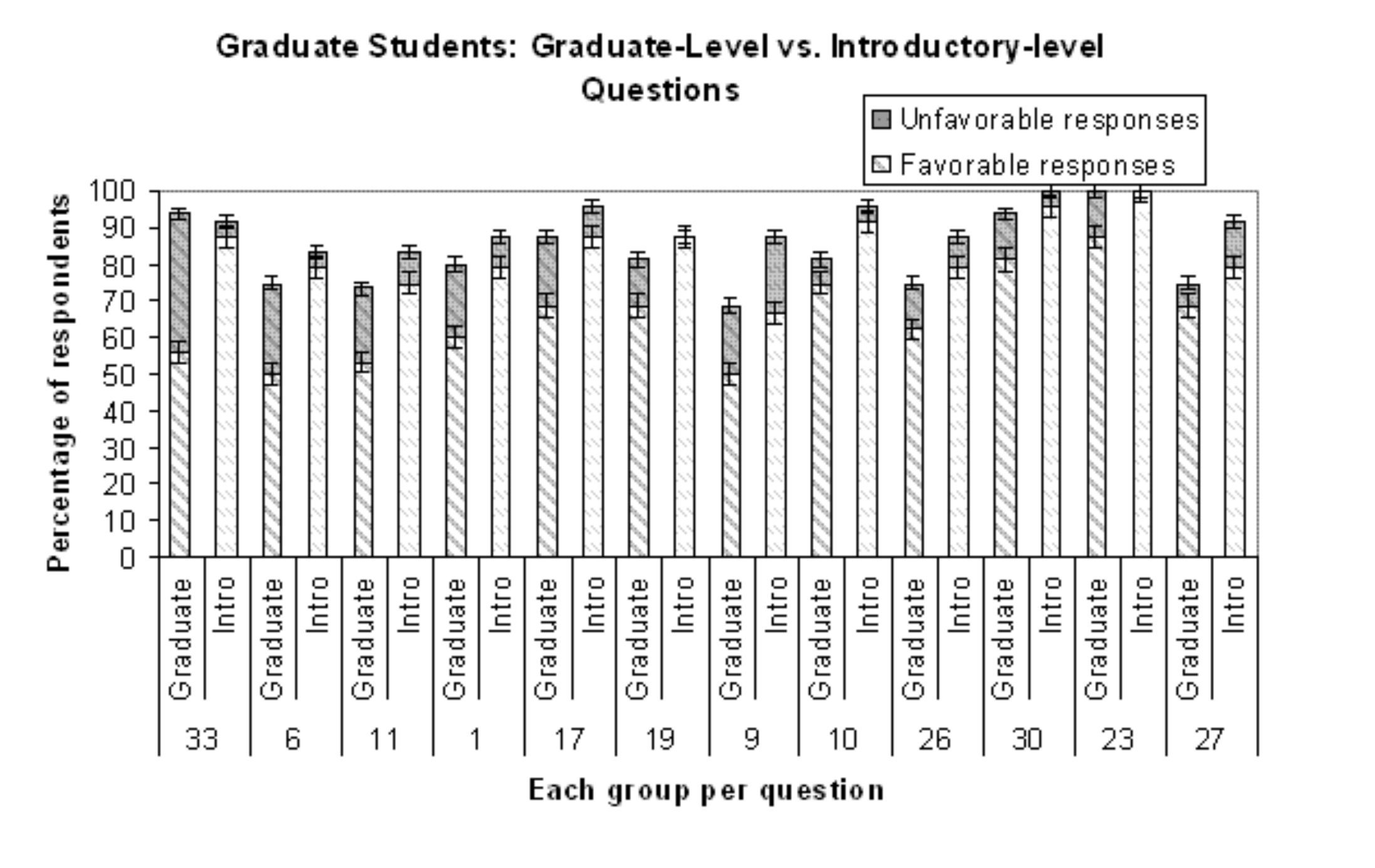}
\caption{Comparison of graduate students' survey responses to 12 selected questions when considering introductory-level problem solving and 
graduate-level problem solving. The order of the questions in the histogram is such that the difference between the introductory-level problem 
solving (``Intro" in the figure) and graduate-level problem solving (``Graduate" in the figure) is largest for the first question (question (33)) 
and second largest for the second question (question (6)) etc. Error bars shown here (and in all other figures) are the standard errors. The 
responses to these questions are more favorable for introductory level problem solving than for graduate-level problem solving.
The neutral percent responses can be found by subtracting from 100, the percentage of favorable and unfavorable responses.}
\end{figure}

\begin{figure}[h!]
  \includegraphics[height=.3\textheight]{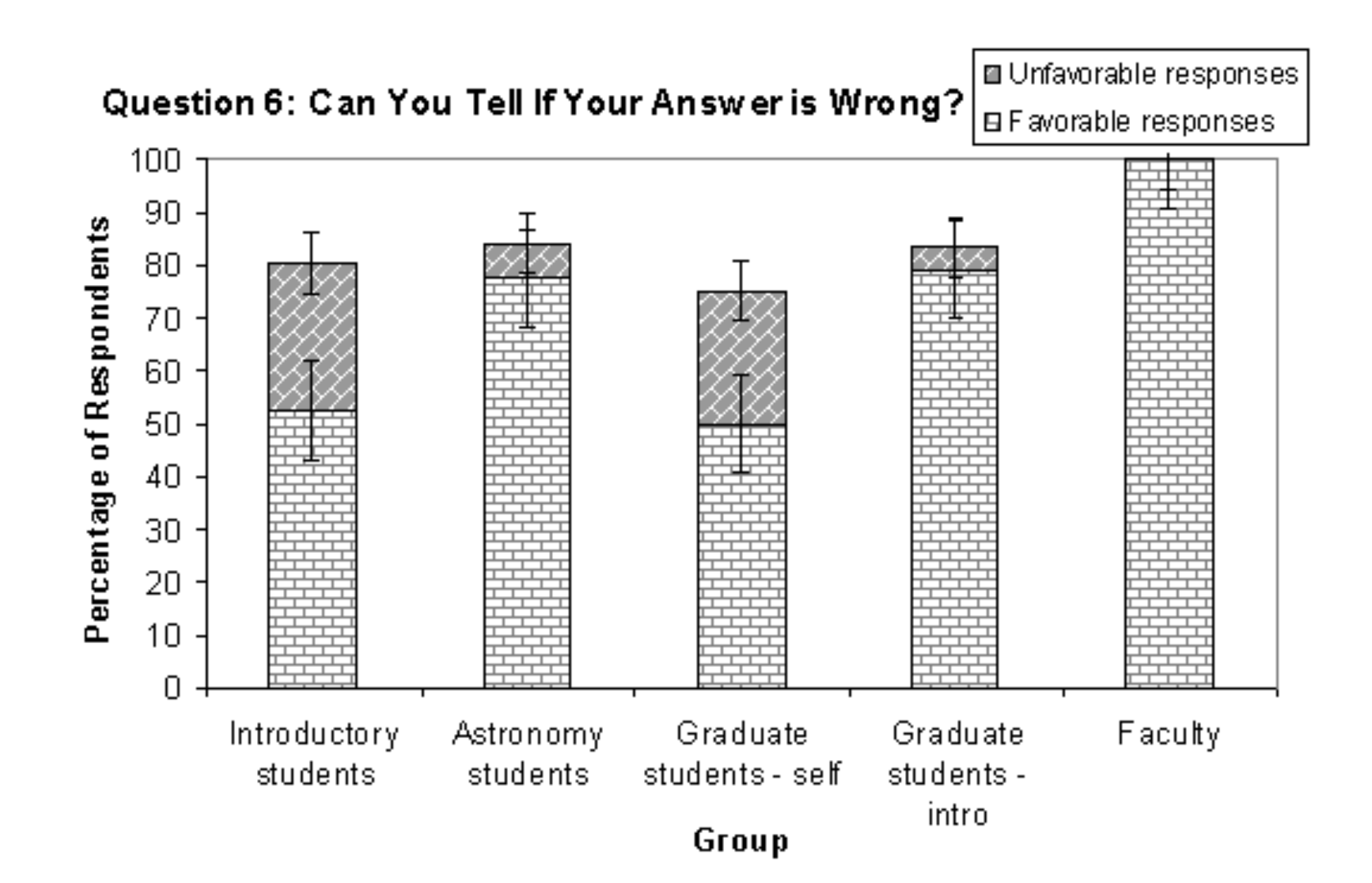}
\caption{Histogram showing favorable (agree) and unfavorable (disagree) responses for survey question (6).
The histogram shows that faculty were always aware of when they were wrong in problem solving but other respondents were less certain.
Only about $50\%$ of graduate students could tell that their answers were wrong in graduate level problem solving.
Graduate students-self and Graduate Students-intro refer to Graduate students' response for graduate level problem solving and introductory level problem solving, respectively.
}
\end{figure}

\begin{figure}[h!]
  \includegraphics[height=.3\textheight]{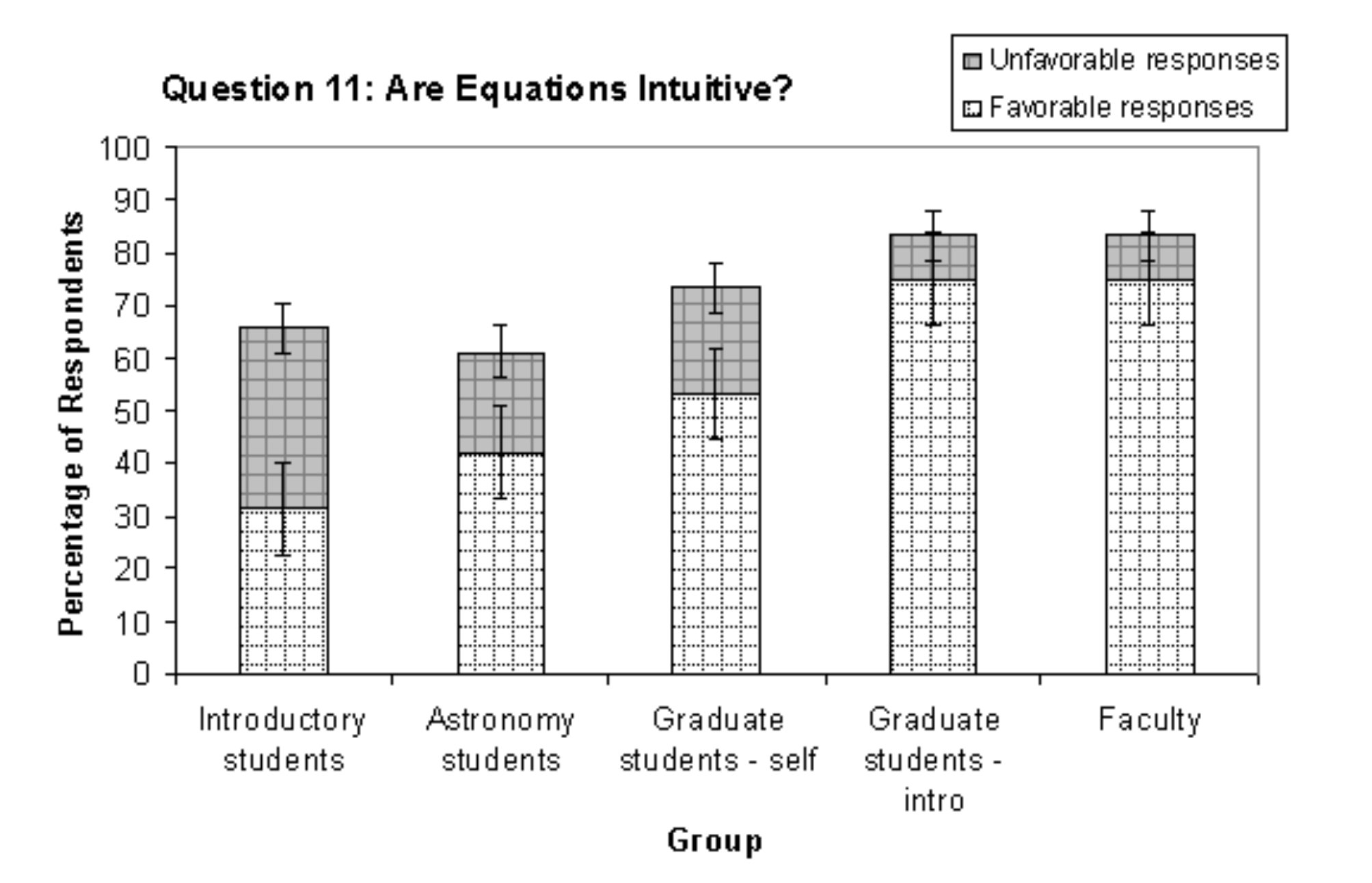}
\caption{Histogram showing favorable (disagree) and unfavorable (agree) responses for survey question (11).
The histogram shows that many more graduate students disagreed that they routinely use equations to calculate answers even if they are non-intuitive
for introductory level problem solving than graduate level problem solving. Almost equal percentage of introductory physics students agreed and
disagreed with the statement.}
\end{figure}

\begin{figure}[h!]
  \includegraphics[height=.3\textheight]{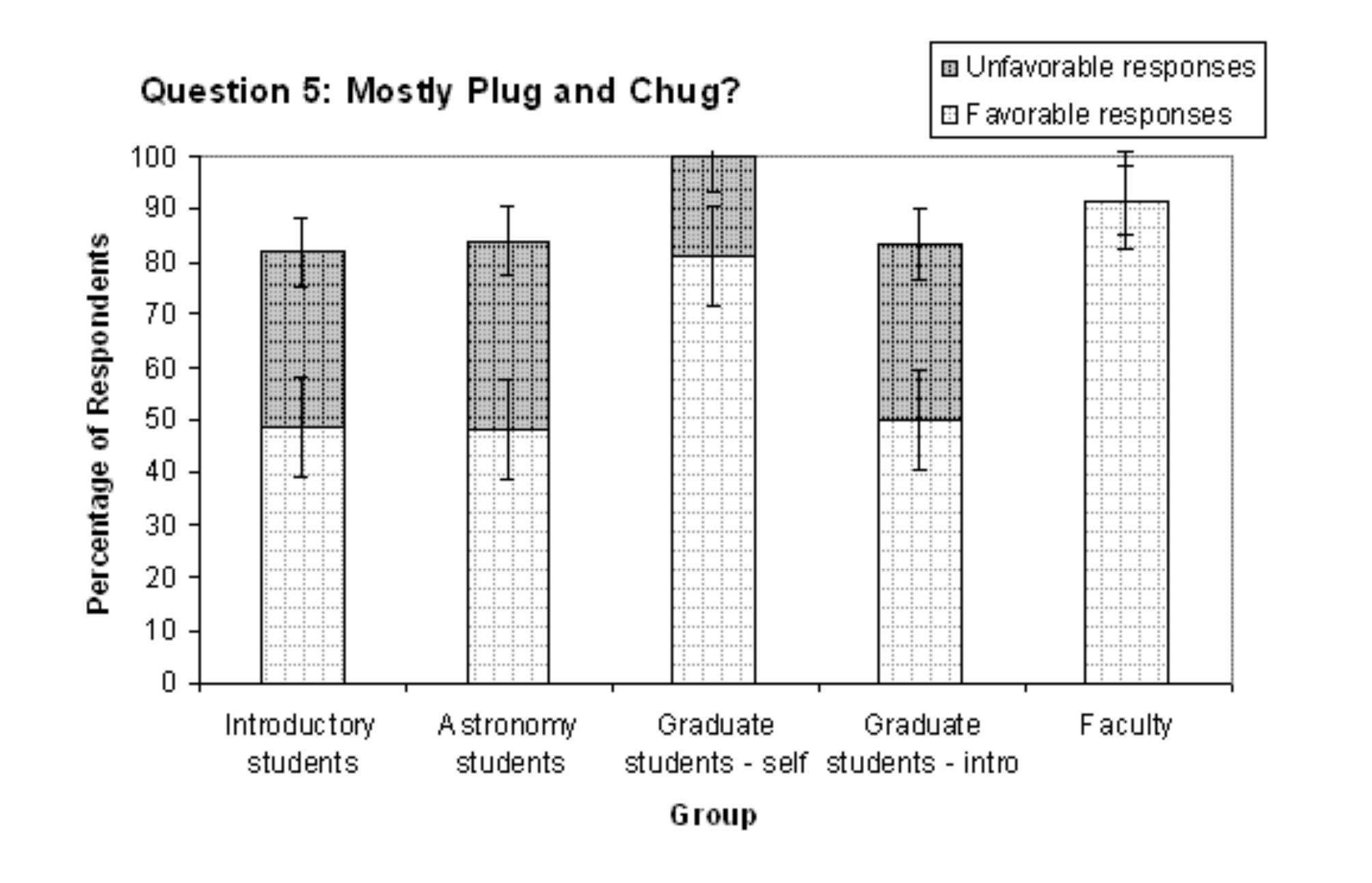}
\caption{Histogram showing favorable (disagree) and unfavorable (agree) responses for survey question (5) about whether problem solving in physics is mainly an exercise in finding the right formula.
The histogram shows that a large percentage of non-faculty respondents from all groups agreed with the statement or were neutral.
}
\end{figure}

\begin{figure}[h!]
  \includegraphics[height=.3\textheight]{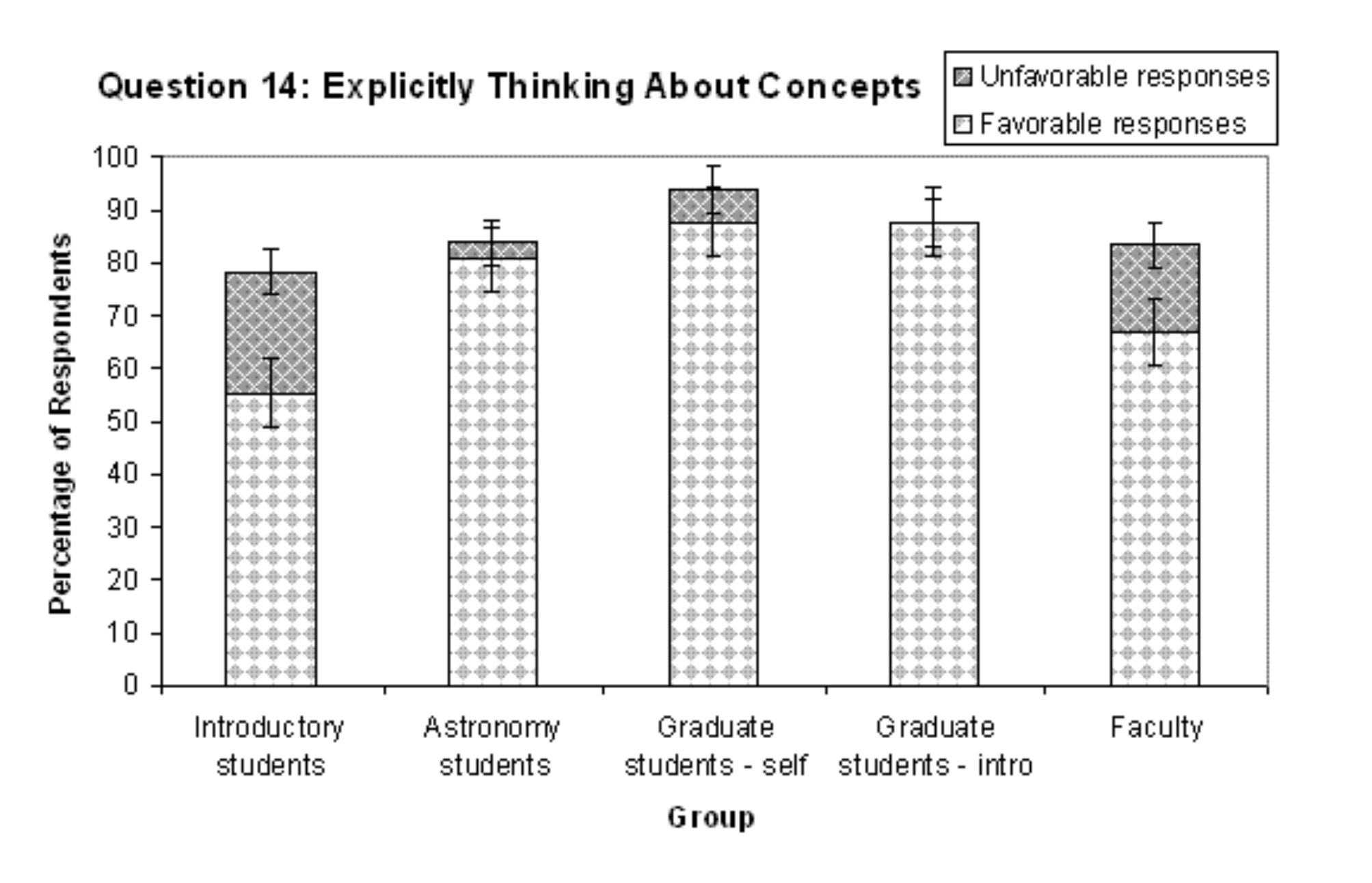}
\caption{Histogram showing favorable (agree) and unfavorable (disagree) responses for survey question (14).
While the trend in the figure from introductory students to faculty may appear to be inconsistent with expectations,
some faculty in individual discussions noted that
they do not explicitly think about concepts that underlie the problem ``while" solving problems because the concepts have become obvious to them.
The introductory students often do not think about concepts because they believe in a plug and chug approach to problem solving in
physics.}
\end{figure}

\begin{figure}[h!]
  \includegraphics[height=.3\textheight]{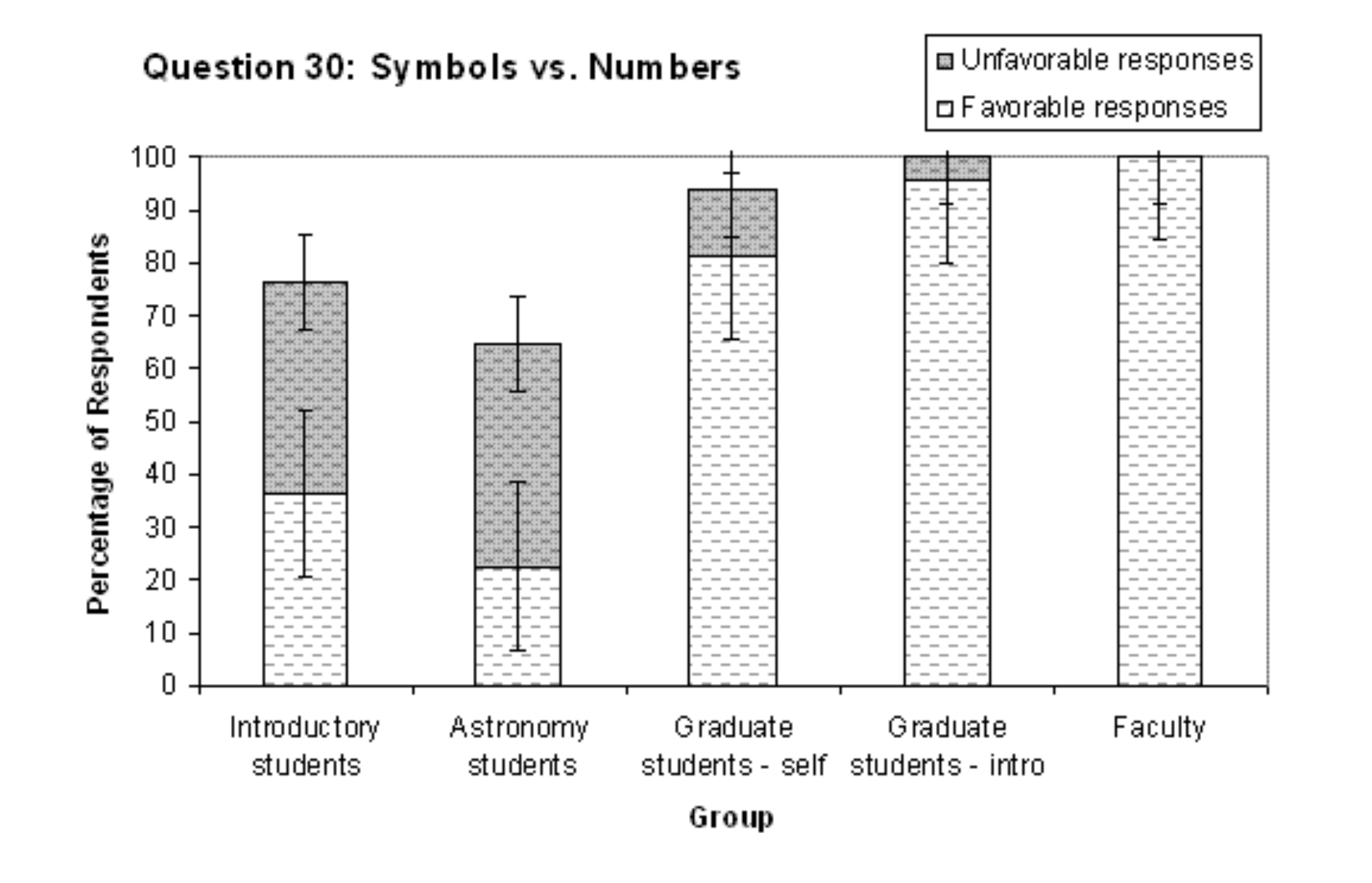}
\caption{Histogram showing favorable (disagree) and unfavorable (agree) responses for survey question (30).
The histogram shows that faculty and graduate students did not believe that it is more difficult to solve a physics problem with symbols than 
solving an identical problem with numerical answer but introductory physics and astronomy students often did.}
\end{figure}

\begin{figure}[h!]
  \includegraphics[height=.3\textheight]{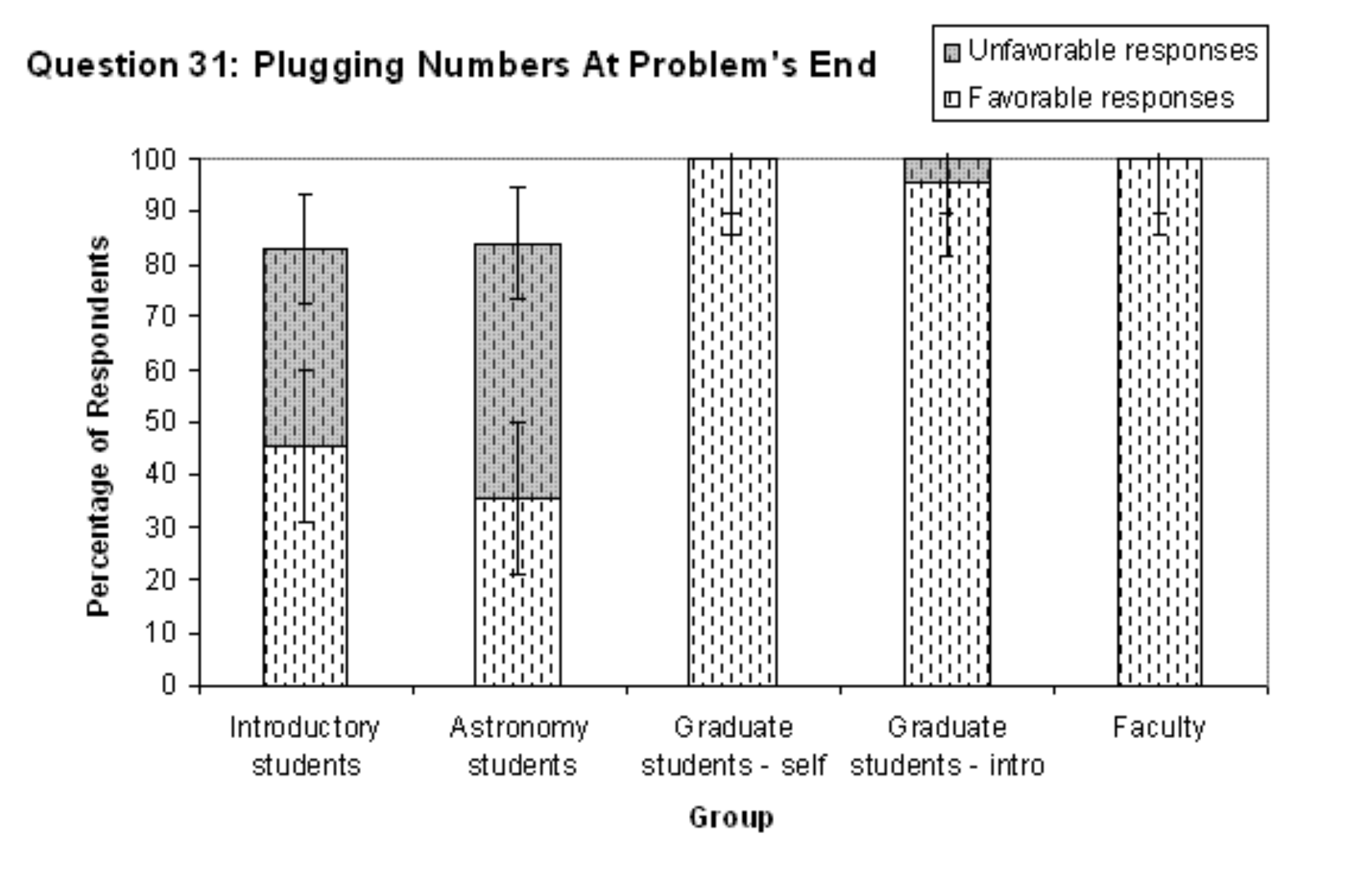}
\caption{Histogram showing favorable (agree) and unfavorable (disagree) responses for survey question (31).
The histogram shows that faculty and graduate students preferred to solve a problem symbolically first and only plug in the numbers at the
very end but less than half of the introductory physics and astronomy students agreed with them.}
\end{figure}

\begin{figure}[h!]
  \includegraphics[height=.3\textheight]{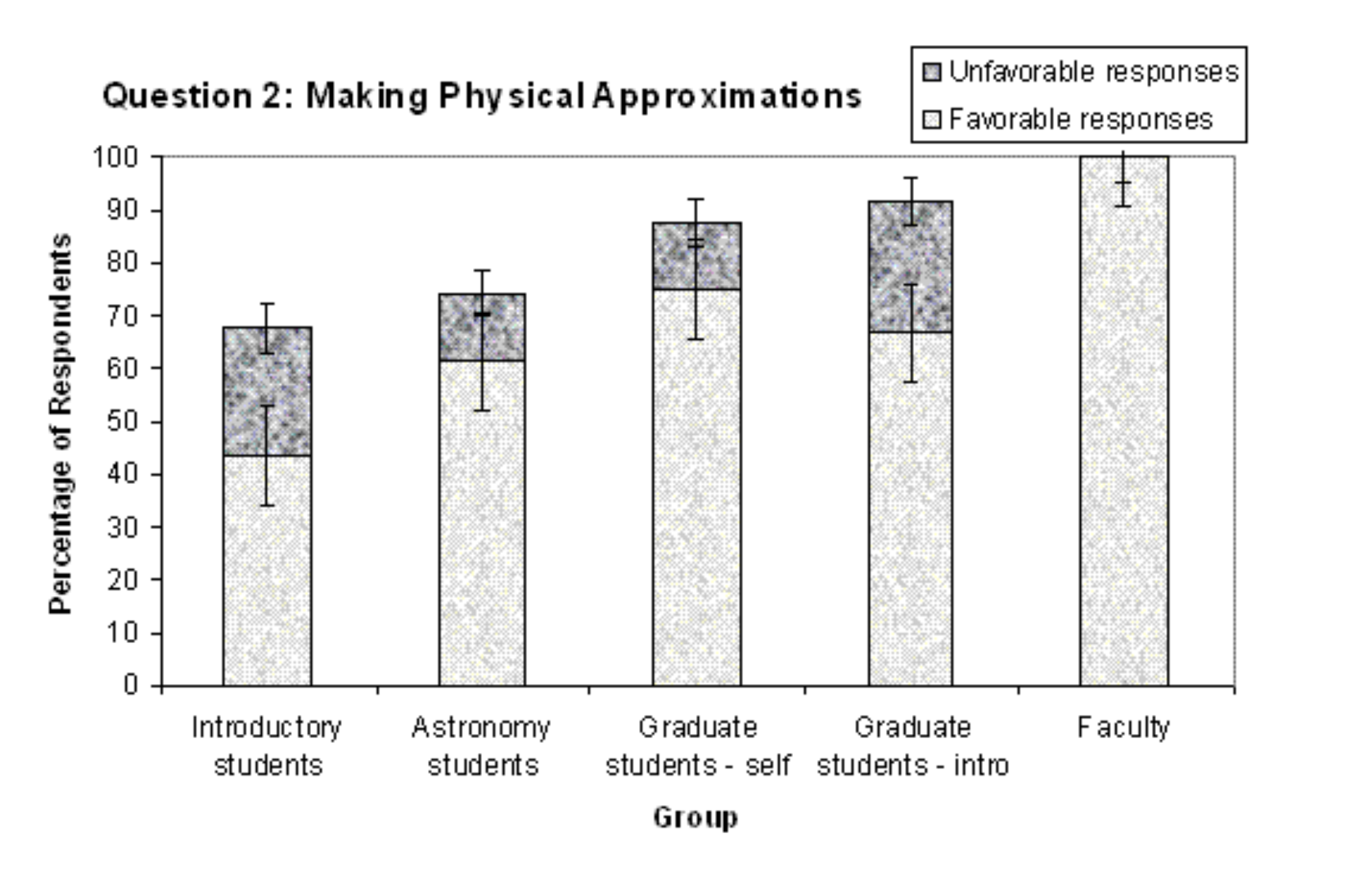}
\caption{Histogram showing favorable (agree) and unfavorable (disagree) responses for survey question (2).
The histogram shows that all faculty agreed that they often make approximations about the physical world
but other respondents, including physics graduate students, were not always in agreement.}
\end{figure}

\begin{figure}[h!]
  \includegraphics[height=.3\textheight]{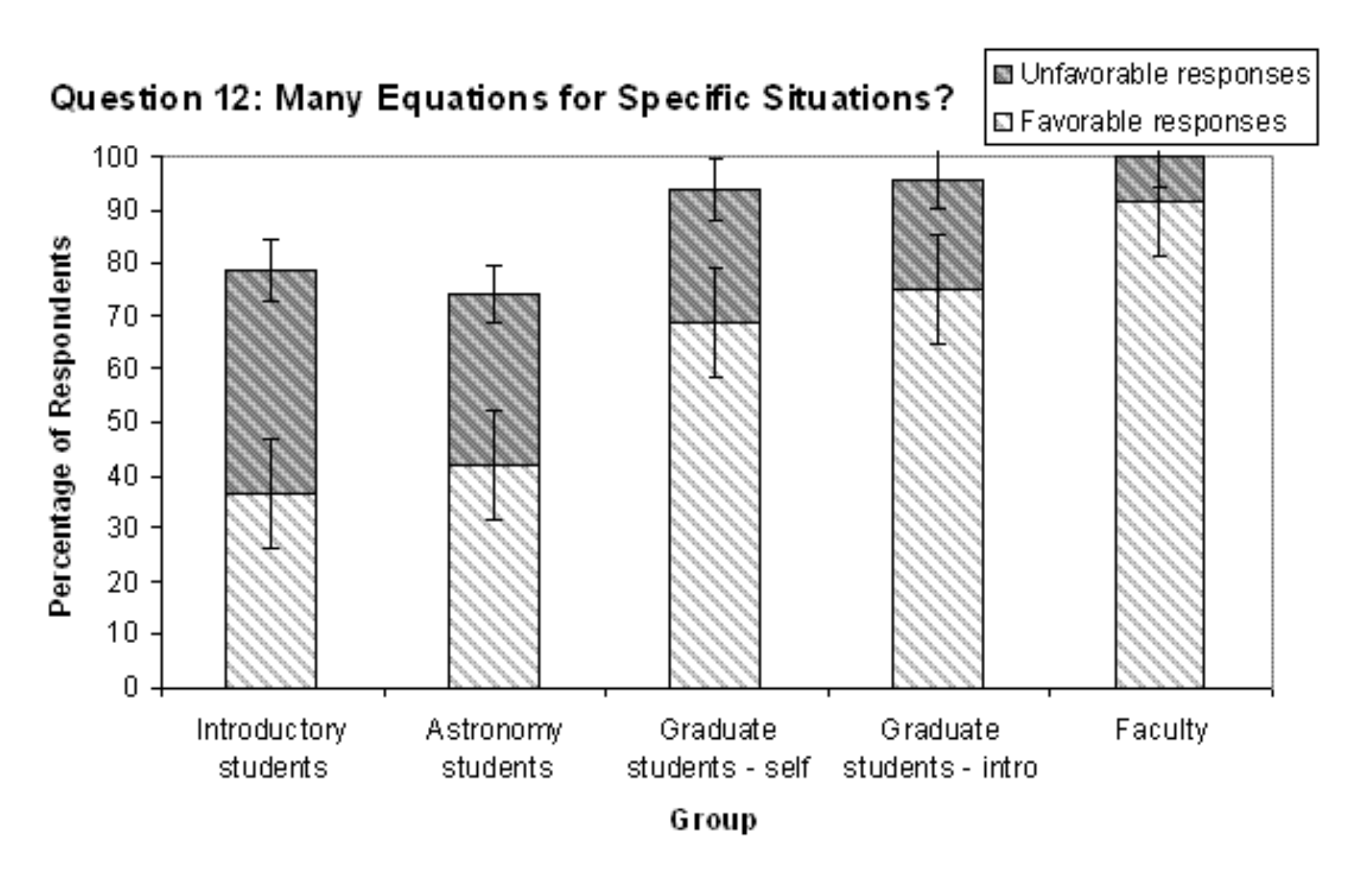}
\caption{Histogram showing favorable (disagree) and unfavorable (agree) responses for survey question (12) about whether physics involves many equations each of which
applies primarily to a specific situation. As we go from the introductory physics and astronomy students to faculty, the disagreement with
the statement (favorable response) increases.}
\end{figure}

\begin{figure}[h!]
  \includegraphics[height=.3\textheight]{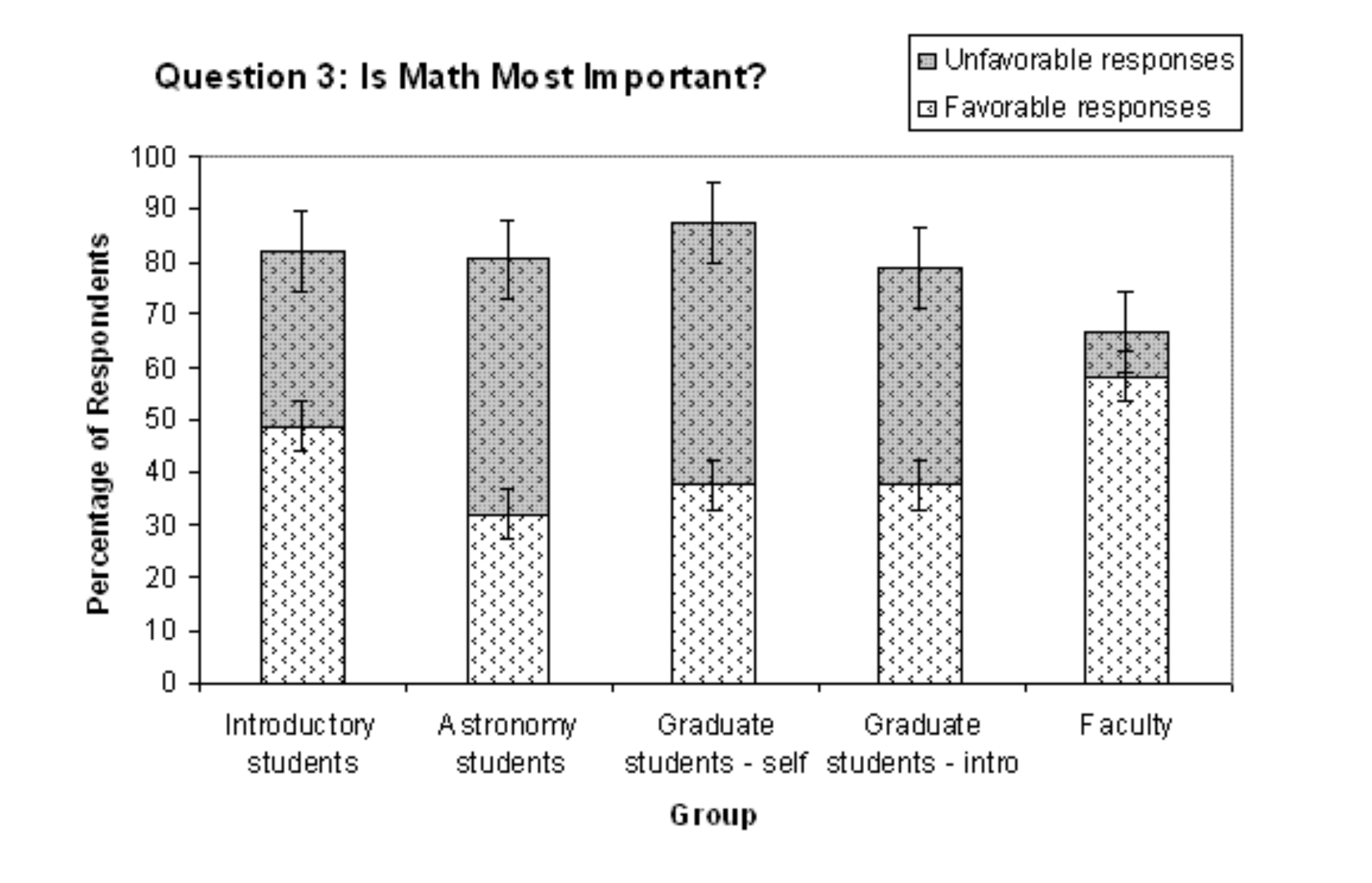}
\caption{Histogram showing favorable (disagree) and unfavorable (agree) responses for survey question (3) about whether mathematics is the most important part of the problem solving process. The histogram shows that a large percentage of non-faculty respondents from all groups agreed with the statement (unfavorable). 
}
\end{figure}

\begin{figure}[h!]
  \includegraphics[height=.3\textheight]{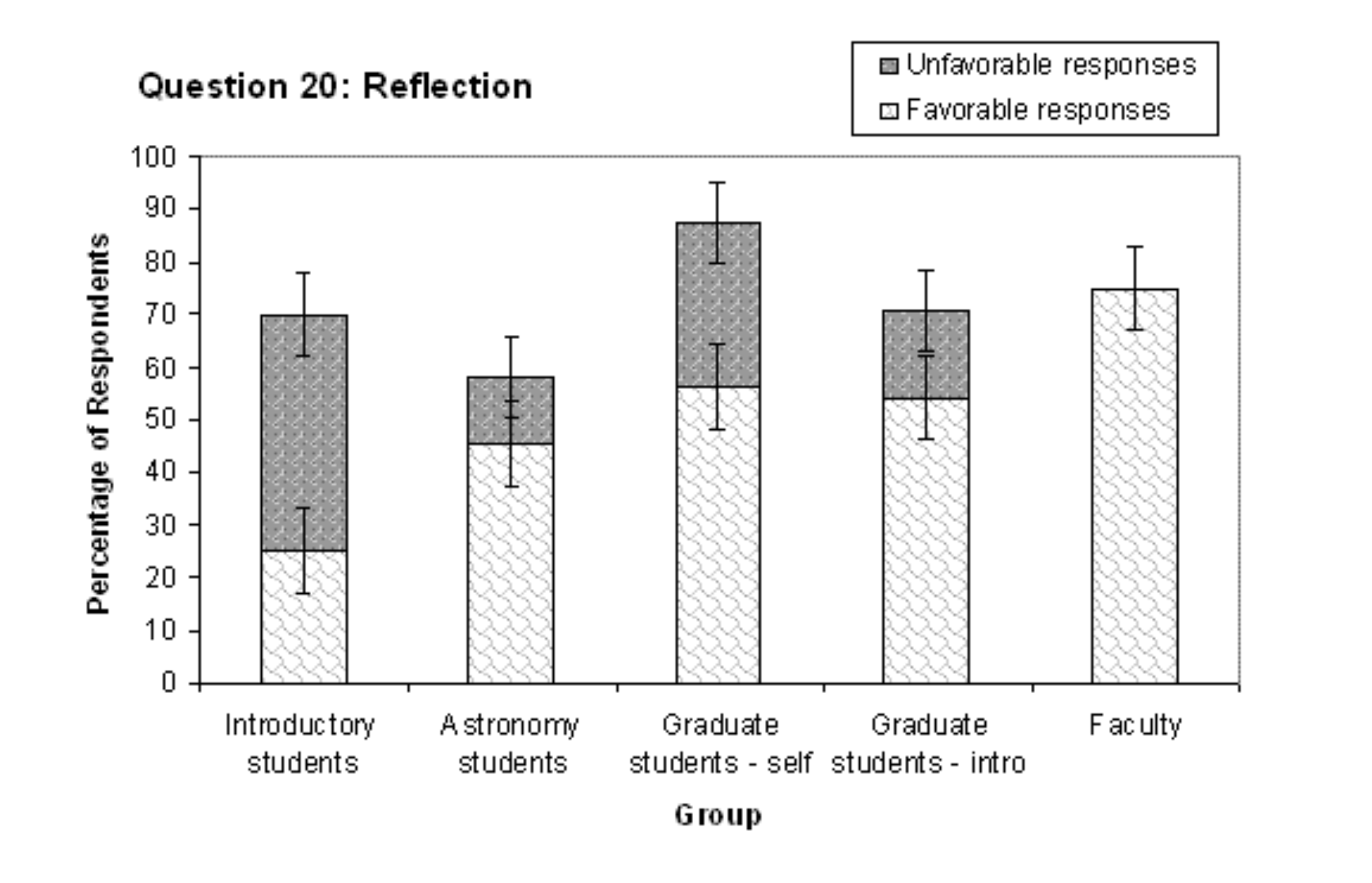}
\caption{Histogram showing favorable (agree) and unfavorable (disagree) responses for survey question (20).
The histogram shows that none of the groups had $80\%$ individuals who agreed that they take the time to reflect and learn from the problem 
solutions after solving problems but the reasons for the lack of reflection varied across different groups.}
\end{figure}

\pagebreak

\end{document}